\documentclass[11pt,onecolumn,amsmath,amssymb,aps,reprint,superscriptaddress,showkeys,pre]{revtex4-2}

\usepackage{graphicx}
\usepackage{bm}%
\usepackage[colorlinks=true,allcolors=blue]{hyperref}%
\usepackage{cleveref}
\usepackage{color}
\usepackage[normalem]{ulem} %
\usepackage{textcomp} %
\usepackage{blindtext}
\usepackage{multirow,booktabs} %

\usepackage{fancyhdr}
\makeatletter
\renewcommand{\figurename}{\textbf{Figure}}
\renewcommand*{\fnum@figure}{{\normalfont\bfseries\figurename~\thefigure}}
\renewcommand*{\@caption@fignum@sep}{\textbf{: }}
\makeatother

\usepackage{siunitx}

\newcommand{\normal}{\mathbf{\hat{n}}}

\newcommand{\chg}[2][black]{{\color{#1}#2}}
\newcommand\rem[1]{}   %
\newcommand\todo[2][black]{{\color{#1}#2}}

\definecolor{jobcolor}{rgb}{0,0,0}%
\def\ethics#1{{\vskip5.5pt\noindent \textcolor{jobcolor}{\fontsize{9}{11}\selectfont Ethics.}\fontsize{8}{11}\selectfont\enskip #1}}
\def\dataccess#1{{\vskip5.5pt\noindent \textcolor{jobcolor}{\fontsize{9}{11}\selectfont Data Accessibility.}\fontsize{8}{11}\selectfont\enskip #1}}
\def\aucontribute#1{{\vskip5.5pt\noindent \textcolor{jobcolor}{\fontsize{9}{11}\selectfont Authors' Contributions.}\fontsize{8}{11}\selectfont\enskip #1}}
\def\competing#1{{\vskip5.5pt\noindent \textcolor{jobcolor}{\fontsize{9}{11}\selectfont Competing Interests.}\fontsize{8}{11}\selectfont\enskip #1}}
\def\funding#1{{\vskip5.5pt\noindent \textcolor{jobcolor}{\fontsize{9}{11}\selectfont Funding.}\fontsize{8}{11}\selectfont\enskip #1}}
\def\ack#1{{\vskip5.5pt\noindent \textcolor{jobcolor}{\fontsize{9}{11}\selectfont Acknowledgements.}\fontsize{8}{11}\selectfont\enskip #1}}

\begin{document}

\linespread{1.15}

\title{A functional shunt in the umbilical cord: the role of coiling in solute and heat transfer}

\author{Tianran~Wan}
\affiliation{Department of Mathematics, University of Manchester, Manchester M13 9PL, UK}
\affiliation{Bioinformatics Institute, A*STAR, 138671, Singapore}
\author{Edward~D.~Johnstone}
\affiliation{Maternal and Fetal Health Research Centre, University of Manchester, Manchester, M13 9WL, UK}
\author{Shier~Nee~Saw}
\affiliation{Department of Artificial Intelligence, Universiti Malaya, Kuala Lumpur, 50603, Malaysia}
\author{Oliver~E.~Jensen}
\affiliation{Department of Mathematics, University of Manchester, Manchester M13 9PL, UK}
\author{Igor~L.~Chernyavsky}
\email[Corresponding author: ]{igor.chernyavsky@manchester.ac.uk}
\affiliation{Department of Mathematics, University of Manchester, Manchester M13 9PL, UK}
\affiliation{Maternal and Fetal Health Research Centre, University of Manchester, Manchester, M13 9WL, UK}

\keywords{umbilical cord, vascular configuration, helicity, oxygenation, diffusive solute transport, heat exchange}

\begin{abstract}\bigskip
 The umbilical cord plays a critical role in delivering nutrients and oxygen from the placenta to the fetus through the umbilical vein, while the two umbilical arteries carry deoxygenated blood with waste products back to the placenta. Although solute exchange in the placenta has been extensively studied, exchange within the cord tissue has not been investigated. Here, we explore the hypothesis that the coiled structure of the umbilical cord could strengthen diffusive coupling between the arteries and the vein, resulting in a functional shunt. We calculate the diffusion of solutes, such as oxygen, and heat in the umbilical cord to quantify how this shunt is affected by vascular configuration within the cord. We demonstrate that the shunt is enhanced by coiling and vessel proximity. Furthermore, our model predicts that typical vascular configurations of the human cord tend to minimise  shunting, which could otherwise disrupt thermal regulation of the fetus. We also show that the exchange, amplified by coiling, can provide additional oxygen supply to the cord tissue surrounding the umbilical vessels.\\[1em] 
\end{abstract}
\maketitle
\onecolumngrid
\vspace{-2em}

\section{Introduction}
    During pregnancy, fetal development depends on the umbilical cord for delivering nutrients and oxygen from the placenta to the fetus through the umbilical vein (UV) and transferring waste products back through the umbilical arteries (UA), before they are discarded through the maternal circulatory system. Because the umbilical cord plays such a vital role during pregnancy, disruption to blood flow or abnormality in the structure of the cord could significantly affect the growth of the fetus \cite{Tantbirojn2009, Makarchuk2023}.
    
\begin{figure}    
    \centering
    \includegraphics[width=0.8\textwidth]{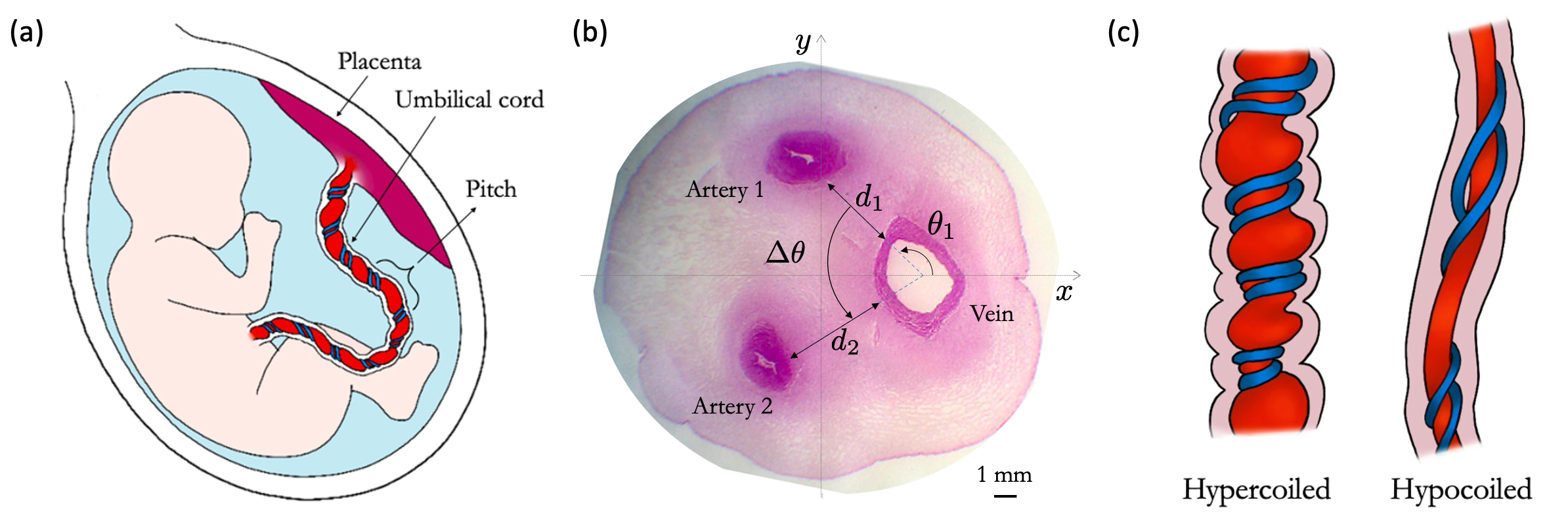}         
    \caption{\textbf{The vascular structure of the umbilical cord.} (a) A schematic of the helical structure of the umbilical cord that connects the fetus to the placenta; the pitch $2\pi/\Omega$ represents the average length of a coil relative to the radius of the cord; $\Omega$ is the dimensionless helicity parameter (proportional to the UCI). (b) A histological image of the cross-section of a healthy human cord (adapted from \cite{Blanco2011} under CC BY-NC 4.0); the vascular cord configuration is characterised by the angle between the umbilical vein and the first artery ($\theta_1$, measured relative to the $x$-axis that passes through the centre of the cord and the centre of the vein), the angle between the two arteries $\Delta\theta$, and the artery--vein separation distances $d_1,\, d_2$ normalised by the cord radius. (c) Schematic illustrations of hypercoiled and hypocoiled umbilical cords.}
    \label{fig:fetus}       
\end{figure}

    The umbilical vessels are characterised by their helical geometry (Fig.~\ref{fig:fetus}). The umbilical coiling index (UCI), defined as the ratio of the number of coils to the length of the cord, or the inverse helical pitch (Fig.~\ref{fig:fetus}a), is commonly used in clinical practice to quantify the coiled geometry \cite{Strong1994}. 
    In normal pregnancies, the mean UCI is approximately 0.17 coils/cm \cite{Dijk2002}; cords with an UCI below the 10th centile (ca. 0.07 coils/cm) are classified as hypocoiled, while cords with an UCI above the 90th centile (ca. 0.3 coils/cm) are classified as hypercoiled (Fig.~\ref{fig:fetus}c). Hypercoiling is associated with an increased incidence of fetal growth restriction (FGR), fetal heart deceleration during delivery, vascular thrombosis and cord stenosis \cite{Machin2000}. Hypocoiling is associated with a higher incidence of fetal demise, fetal distress at delivery and chromosomal abnormalities \cite{Strong1993}.

 Given the association between abnormal coiling and adverse pregnancy outcomes, it is important to explore how features of helical cord structure, such as coiling and vascular configuration (Fig.~\ref{fig:fetus}b), may affect umbilical cord function. 
 In the umbilical cord, many solutes come from the placenta, where exchange between fetal and maternal blood takes place. While placental exchange has been widely investigated \cite{Jensen2019,Clark_2015, Pearce2016, PlitmanMayo2016}, and a few studies have analysed blood flow in umbilical vessels \cite{Waters01, Wilke2018, Kaplan2009, Saw2017}, to the best of our knowledge, solute exchange between the umbilical vessels remains unexplored.

In addition to its role in solute transport, the cord also contributes to fetal thermoregulation, which is immature compared to adult physiology, as the fetus lacks the ability to sweat or shiver. Instead, the fetus relies on heat exchange through the feto-placental circulation to maintain an appropriate temperature. Extreme heat exposure or maternal hyperthermia during pregnancy
can overwhelm the fetus's limited ability to regulate its own temperature, potentially leading to complications such as preterm birth, low birth weight and congenital anomalies \cite{Syed2022}. A recent computational study by Kasiteropoulou \emph{et al.}~\cite{Kasiteropoulou2020} indicated the importance of umbilical coiling in thermal regulation; however, the overall role of cord structure in feto-maternal heat exchange remains an open question.

Finally, in humans, the cord's connective tissue, known as Wharton's jelly (WJ), has no extra vasculature besides the umbilical vessels \cite{Benirschke2012}. In other mammals, several smaller vessels have been found through histological studies \cite{Benirschke2007}. It has been hypothesised that they may act as \emph{vasa vasorum}, smaller vessels that supply the walls of larger blood vessels. Given the coiled nature of the human umbilical vessels, an enhanced exchange could also affect oxygen availability for vessel walls and cells in the extravascular cord tissue. 

The objective of this study is therefore to quantify the diffusive coupling of umbilical vessels in the human cord for different solutes and heat by combining a theoretical model with anatomical data.
Our model moves beyond the coiling index by encapsulating the complexity of three-dimensional helical vascular configurations in a computationally efficient and interpretable manner, allowing us to define key geometric determinants of functional shunting in the cord for a wide range of solutes.
We further contextualise our results for specific structural features of human umbilical cords, based on \emph{ex vivo} histology and \emph{in vivo} ultrasound imaging, and show how typically observed vascular configurations tend to minimise heat exchange between the cord's vessels.
Our model also predicts that cord helicity can enhance oxygen availability to the cord tissue surrounding the umbilical vessels, particularly in the presence of tissue metabolism. The developed methodology provides a building block for future fetal--placental--maternal models, enabling more direct hypothesis testing of structure–function relationships in obstetric pathophysiology, as well as in other complex exchange organs.

\section{Methods}
Our integrative approach is outlined below, starting with a theoretical model that seeks to capture the dominant features of intervascular solute exchange in a coiled umbilical cord.
\subsection{Modelling solute exchange between helical vessels}

\begin{table}
\centering
\begin{tabular}{llllll}
\toprule
\textbf{Parameter}         & \textbf{Notation}\hspace{2em} & \textbf{Value (cm)}\hspace{2em}  & \textbf{Source} \\ 
\midrule
Arterial radius            & $r^*_a$  & $0.06-0.2$  & \cite{Weissman1994} \\
Venous radius              & $r^*_v$  & $0.1-0.4$   &  \cite{Weissman1994} \\ 
Cord radius                & $R^*_c$  & $0.2-1\,^{\dagger}$   & \cite{Raio1999} \\  
Venous helical radius      & $R^*_v$  & $0 - 0.8$   & \cite{Kaplan2009} \\ 
Venous wall thickness      & $h^*_v$  & $0.03 - 0.05$  & \cite{Gayatri2017} \\
Arterial wall thickness    & $h^*_a$  & $0.05 - 0.08$  & \cite{Gayatri2017} \\
Cord pitch                 & $2\pi/\Omega^*$  & $3 - 14$ &  \cite{Laat2005} \\ 
Cord length                & $L^*$    & $50-100$ & \cite{Benirschke2012} \\
Artery--vein distance \hspace{2em} & $d^*$    & $0.04 - 0.4$ &  Estimated \\
\bottomrule
\end{tabular}
\caption{Key %
geometric parameters of the model. Stars denote dimensional quantities. The pitch was calculated as the inverse umbilical coiling index (UCI)~\cite{Laat2005}, and the artery--vein distance was estimated as described in Supplement~S6.\linebreak
$^{\dagger}$ All the dimensionless lengths in the model are scaled on $R^*_c = 1$\,cm.}
\label{tab:par}
\end{table}

We model the umbilical cord as a straight circular cylinder that hosts three helical vessels (one vein and two arteries). Each vessel has a circular cross-section in the plane orthogonal to the axis $z$ of the cylinder and a constant (and common) pitch $2\pi/\Omega$, defined as the length of one complete helical turn parallel to the $z$-axis of the cord, measured relative to the cord's radius (see Fig.~\ref{fig:fetus}a). The dimensionless helicity parameter $\Omega$ is proportional to the UCI. The vascular configuration in the cross-section of a cord (Fig.~\ref{fig:fetus}b) can be characterised by the relative distance $d$ between the umbilical arteries and the vein (scaled by the cord's radius) and the angles between the vein and each of the two arteries $\theta_1$ and $\theta_2$ (measured with respect to an axis connecting the centre of the cord to the centre of the vein). For simplicity, we assume that both umbilical arteries are of the same size and are the same distance $d$ from the vein (so that $d_1=d_2=d$ in Fig.~\ref{fig:fetus}b).

We model the transfer of solutes (such as oxygen) and heat in the cord tissue $\mathcal{I}_\text{t}$ by a steady diffusion-uptake process and assume the cross-sectional solute concentration to be approximately uniform in each umbilical vessel.
While blood flow in the umbilical vein is approximately steady, with weak fluctuations \cite{Rubin22}, arterial flow is pulsatile \cite{Waters01}. 
However, the diffusion (with tissue diffusivity $D^*_t$) of oxygen and heat between the vessels is much slower ($(d^*)^2/D^*_t \sim 10 - 10^3$\,s, for $d^* \sim 1$\,mm; see Table~\ref{tab:par} and Supplement) compared to the typical fetal heart-rate timescale
($10^{-1} - 1$\,s), justifying the steady-transport approximation. %
We also neglect transmural flow in the cord tissue \cite{TuanMu_etal20} and possible differences in solute concentration between the umbilical arterial blood and the amniotic fluid. The governing equations (see Supplement \todo{S1} for more details) in dimensionless variables describing solute transport in the umbilical tissue are given by
\begin{subequations}
\label{eq:3D-transport}
\begin{align}
    \nabla^2 C &= \alpha \left(C + 1/\beta\right) \quad \text{in }\; \mathcal{I}_\text{t}\,, \label{eq:3difcart}\\   
    \eta\, C + (\normal \cdot \nabla C)  &= 0 \quad \text{on }\; \mathcal{I}_\text{c}\,,\label{eq:outbc}\\
    \normal \cdot \nabla C&= 0 \quad \text{on }\; \mathcal{I}_\text{f,\,p}\,, \label{eq:fpside}\\
    C &= 0 \quad \text{on }\; \mathcal{I}_\text{a$_i$}\; \text{for }\; i=1,2\,,\quad 
    \text{and}\quad C = 1\quad \text{on }\; \mathcal{I}_\text{v}\,. \label{eq:3DCi} 
\end{align}
\end{subequations} 
Here $C(x,y,z)$ is the dimensionless solute concentration (or temperature) normalised by the vein-artery concentration difference $C_v-C_a$ and centred at the arterial concentration $C_a$, so that the dimensional concentration is $C_a +(C_v-C_a)C$. In \eqref{eq:3difcart}, $\alpha$ is a dimensionless metabolic rate parameter (assumed zero for all the solutes but oxygen), and $\beta=(C_v-C_a)/C_a$ is the relative solute concentration difference between the umbilical artery (UA) and the umbilical vein (UV) (see Supplement \todo{S1}). Exchange at the interface $\mathcal{I}_\text{c}$ between the umbilical cord and the amniotic fluid is characterised by the amniotic exchange parameter $\eta$ in \eqref{eq:outbc}; when $\eta=0$ no solute is lost through outer cord's surface; $\eta \to \infty$ in \eqref{eq:outbc} enforces $C=0$ at the cord--amniotic interface.
We impose no-flux conditions in \eqref{eq:fpside} at the tissue cross-sections bounding the fetal and placental ends of the umbilical cord respectively ($\mathcal{I}_\text{f,\,p}$); here $\normal$ denotes a unit normal pointing outside the domain occupied by the cord tissue $\mathcal{I}_\text{t}$. Finally, assuming zero axial concentration gradient along the cord, we set fixed solute concentrations on the surface of the umbilical arteries $\mathcal{I}_\text{a$_1$}$, $\mathcal{I}_\text{a$_2$}$ and the vein $\mathcal{I}_\text{v}$ in \eqref{eq:3DCi}, so that we can focus attention on intervascular exchange through cord tissue, accounting for its internal helical structure.

We reduce the three-dimensional problem \eqref{eq:3D-transport} to two dimensions by exploiting the helical coordinate transformation $X = x\cos{\left(\Omega z\right)} + y \sin{\left(\Omega z\right)}$, 
$Y = -x\sin{\left(\Omega z\right)} + y \cos{\left(\Omega z\right)}$, $Z=z$, where $z$ measures the distance along the cord's axis (see Supplement \todo{S1} for more details). This transformation allows us to study the three-dimensional effects of helicity in the cross-sectional plane of the cord by solving for $C(X,Y)$, the field that is invariant along the length of the cord relative to a rotating coordinate system.
\chg{We assume that all vessels share the same pitch and that each vessel centreline maintains a fixed distance from the (straight) cord centreline.}
The dimensionless solute flux $N$ per unit length of the cord delivered from the UV to the two UAs and cord tissue, 
and the solute flux per unit length taken up by cord tissue are then given, respectively, by 
\begin{equation}
    N(\Omega,\alpha,\beta,\eta) = \int_{\partial\mathcal{I}_\text{v} \,\cap\,\mathcal{P}} 
      \normal \cdot \left( \nabla_{\perp} C  + \Omega^2 \mathbf{H} \right) \mathop{}\mathrm{d}s \quad\;\text{and}\;\quad 
    N_\text{u} = \int_{\mathcal{I}_\text{t} \,\cap\,\mathcal{P}} \!\alpha\left(C + 1/\beta\right)\!\mathop{}\mathrm{d}A\,,
    \label{eq:Nfluxes}
\end{equation}
where $\nabla_\perp \equiv [\partial_X,\,\partial_Y,\,0]$ is the 2D gradient operator, and $\mathbf{H} = (Y\,C_X - X\,C_Y)\,[Y,\,-X,\,0]$ is a flux component arising due to cord helicity (see Supplement \todo{S2} for further details). The integrals in \eqref{eq:Nfluxes} are taken in a plane $\mathcal{P}$, in which $z=\text{constant}$.  The solute flux $N(\Omega,\alpha,\beta,\eta)$ 
in transformed coordinates shows explicitly the rotational effects of helicity. The helicity factor $\Omega^2$ controls the strength of the rotational effects while $\mathbf{H}$ represents how the concentration gradients twist in space. This term shows that concentration not only diffuses radially but also experiences angular distortions relative to the vessels. We explore below the effects of helicity, vessel proximity and the configuration of the vessels within the cord on the solute fluxes. The ratio of a dimensional exchange flux between the umbilical vessels, based on $N$, and the advective flux along the umbilical cord defines a dimensionless Damk\"{o}hler number
\begin{equation}\label{eq:Da}
    \mathrm{Da} = \frac{ D^{*}_t L^{*}\,N(\Omega,\alpha,\beta,\eta)}{B\,Q^{*}}\,,
\end{equation}
where $D^{*}_t$ is the solute diffusivity in tissue, $L^{*}$ is the cord length, $Q^{*}$ is the total umbilical flow rate and $B$ is an advection-facilitation parameter that accounts for haemoglobin binding of certain solutes, such as oxygen; see Supplement \todo{S8} and Table~S3 for more details and estimates of $\mathrm{Da}$ for oxygen and heat exchange.

As the metabolic rate $\alpha$ increases, the region surrounding the vessels' lumens exhibits thin boundary layers and steep concentration gradients.
We exclude the effects of these highly localised boundary layers by calculating the extravascular uptake flux within the cord tissue outside a thin strip of thickness $h$   around each vessel (based on typical values for the thickness of the umbilical vessel walls \cite{Gayatri2017}; see Table~\ref{tab:par} and Supplement). 
The solute distribution and the corresponding integral fluxes \eqref{eq:Nfluxes} are computed via a finite-element solver of COMSOL Multiphysics$^\text{\texttrademark}$ 6.0. 
We validated the use of the 2D field $C(X,Y)$ against direct computationally-expensive solutions of the 3D problem \eqref{eq:3D-transport} (see Supplement \todo{S5} for more details).

\subsection{Image analysis of histology and ultrasound data}

Histology and ultrasound images of human cord cross-sections were taken from published literature \cite{Blanco2011, Thomas2020, Herzog2017, BlancoElices2022, Benirschke2012,Kurita2009,Kurakazu2019, HenanDh2016}. The histology data (stained with hematoxylin and eosin; see Fig.~\ref{fig:fetus}b) corresponds to normal pregnancies, with gestational age varying from 37 to 42 weeks; ultrasound data included cords from the third trimester (see Supplementary Table \todo{S4} for more details on data sources).

A semi-automated Python script (see Supplement \todo{S6}) was used to extract structural metrics from the images,
including effective radii and distances between each umbilical vessel, as well as  the two characteristic angles (Fig.~\ref{fig:fetus}b). All measurements were normalised by the cord's radius such that the scaled cord's radius for each measurement is of unit length. The shape irregularity of each cord was quantified by the circularity parameter $4\pi\mathcal{A}/(\mathcal{P})^2$, where $\mathcal{A}$ is the cross-sectional area and $\mathcal{P}$ is the perimeter of the cord.

\section{Results}

\subsection{Cord coiling and vascular proximity promote solute and heat shunting}\label{sec:coild}

\begin{figure}
    \centering
    \includegraphics[width=0.7\textwidth]{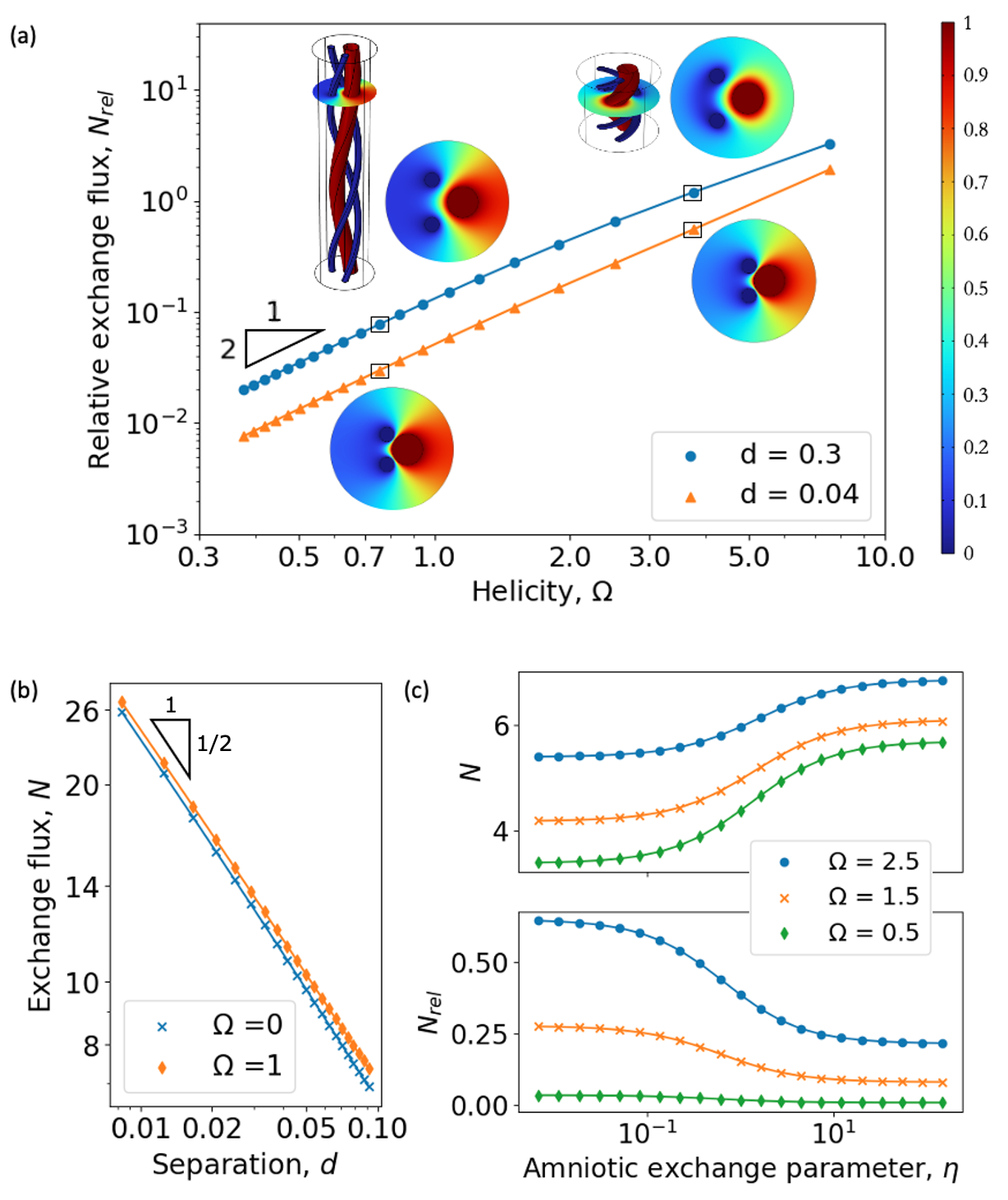}
    \caption{\textbf{The impact of cord helicity and vascular proximity on solute exchange.} (a) Relative exchange flux $N_\mathrm{rel}$ per unit length \eqref{eq:flux_rel_uncoiled} of the umbilical vein vs. cord helicity $\Omega$ for different UA-UV separation distances $d=0.3$ and $d=0.04$, with $\alpha =0,\, \eta=0$. For weakly coiled cords, the solute (heat) exchange amplification scales as $\Omega^2$ (relative to the uncoiled vessels of the same length).  
    Concentration fields for four different vessel configurations and three-dimensional diagrams show a normocoiled ($\Omega$=0.75) and hypercoiled ($\Omega$=3.77) cord, at parameters highlighted with squares on the graph. Cords with $\Omega<0.4$ are hypocoiled while cords with $\Omega>2$ are hypercoiled. The colour bar shows solute concentration in normalised units (venous concentration is unity and arterial concentration is zero). (b) Computed exchange flux per unit length of the vein $N$ for small separation $d$. In the limit of small $d$, the leading-order flux is $\mathcal{O}(d^{-1/2})$. 
    Here $\alpha =0,\, \eta=0$. (c) Exchange flux $N$ and relative exchange flux $N_\mathrm{rel}$ for different values of the exchange parameter $\eta$ in the cord boundary condition \eqref{eq:outbc} for $d = 0.33$. Other geometric parameters in (a-c) were fixed: $R_v = 0.25$, $r_v = 0.25$, $r_a = 0.12$, $\theta_1 = 0.8\pi$, $\theta_2 = 1.2\pi$ (see Supplement for more details). The triangles in (a) and (b) show the relative fold-change in the variables.}
    \label{fig:flux_coil}
\end{figure}

We first explore how the structural parameters affect solute fluxes between the three vessels in a cord. Fig.~\ref{fig:flux_coil}(a) illustrates the difference in exchange flux for helical ($\Omega>0$) and straight ($\Omega=0$) vessel geometries, in terms of the relative exchange flux
\begin{equation}\label{eq:flux_rel_uncoiled}
    N_{rel} =\frac{N(\Omega,\alpha,\beta,\eta)-N(0,\alpha,\beta,\eta)}{N(0,\alpha,\beta,\eta)},
\end{equation}
assuming initially that $\alpha=\eta=0$.
Small $\Omega$ corresponds to weakly coiled cords, while a large $\Omega$ corresponds to a tightly coiled cord. This is shown in the 3D representations of one complete coil for $\Omega=0.75$ and $\Omega=3.77$ (Fig.~\ref{fig:flux_coil}a, insets). 
For large $\Omega$, $N_{rel}$ exceeds unity, implying that the flux can be twice as large as the flux for straight vessels, because tightly coiled cords provide a larger internal surface area for exchange to take place. The concentration field for $\Omega=3.77$, $d=0.3$ (Fig.~\ref{fig:flux_coil}a, inset, top right) varies weakly around the cord perimeter, showing how coiling promotes direct solute exchange between vessels. In contrast, the flux for weakly coiled vessels is similar to the flux for straight vessels. Fig.~\ref{fig:flux_coil}(a) shows that the first correction to the straight vessel flux is of order $\Omega^2$. A weak helicity $\Omega \ll 1$ approximation for flux can therefore be obtained by perturbing the three-dimensional problem \eqref{eq:3difcart} in powers of $\Omega^{2}$. 

Since in the problem for $C(X,Y)$ the helicity term appears as $\Omega^2$, there is no evidence of chirality in the concentration fields in Fig.~\ref{fig:flux_coil}(a). In the Supplement (Fig.~\todo{S8}b,c), we confirm that two configurations with opposite chirality have the same cross-sectional concentration fields. The position of the arteries in the configuration considered in Fig.~\ref{fig:flux_coil}(a) is symmetrical relative to the $x$-axis. This leads to the symmetry $C(X,Y)=C(X,-Y)$. The Supplement (Fig.~\todo{S8}a) shows concentration fields with increasing helicity for asymmetric configurations. Breaking the symmetry introduces distortion in the concentration fields, but the trend in relative flux is the same as in Fig.~\ref{fig:flux_coil}(a).

Another geometric parameter that contributes to %
the exchange of solutes is the separation distance $d$ between vessels. The flux is larger for the umbilical vessels that are closer together because concentration gradients are steeper, as shown 
in Fig.~\ref{fig:flux_coil}(a), colour map inserts. Indeed, Fig.~\ref{fig:flux_coil}(b) illustrates how, as $d$ gets smaller, the flux in the gap between the vessels of $\mathcal{O}(d^{-1/2})$ dominates over the flux outside the gap (see Supplement~\todo{S3} for more details), and the weak helicity correction of $\mathcal{O}(d^{1/2}\,\Omega^{2})$ provides a good approximation ($d = 0.04$; Fig.~\ref{fig:flux_coil}a). On the other hand, for larger $d$, the effect of helicity becomes more dominant, as demonstrated by the larger excess flux $N_{rel}$ \eqref{eq:flux_rel_uncoiled} relative to the flux in the uncoiled geometry.

Fig.~\ref{fig:flux_coil}(c) illustrates the effects of changing the amniotic exchange parameter $\eta$, which controls the outer boundary condition \eqref{eq:outbc}, on the exchange flux. When $\eta \to 0$, the outer boundary is completely isolated, and the concentration gradients in the cord tissue are driven entirely by the interaction between the umbilical vessels, as in Fig.~\ref{fig:flux_coil}(a,b).  
When $\eta \to \infty$, the flux increases  
since the additional sink, the  amniotic fluid, forces steeper concentration gradients within the cord tissue.
For weakly coiled cords, $N$ almost doubles as $\eta \to \infty$. In contrast, the relative difference in exchange flux $N_\mathrm{rel}$ for different $\Omega$ shows that the role of helicity is strongest for $\eta=0$  
($N_\mathrm{rel}(\Omega=2.5)-N_\mathrm{rel}(\Omega=0.5) \approx 0.6$), 
compared to $\eta \to \infty$, when this difference falls ($N_\mathrm{rel}(\Omega=2.5)-N_\mathrm{rel}(\Omega=0.5) \approx 0.2$). We use the no-flux condition ($\eta=0$) for the remainder of this work because the effect of helicity is maximised, providing an upper bound for the relative diffusive coupling $N_{rel}$ between the umbilical vessels in this limit.

For the range of exchange fluxes $N$ predicted in Fig.~\ref{fig:flux_coil}, the characteristic values of the Damk\"ohler number \eqref{eq:Da}, which quantifies the extent of `shunting' between the UV and UA, are $\mathrm{Da}_{\text{O}_2} \sim 10^{-5}$ and $\mathrm{Da}_\text{heat} \sim 10^{-1}$ for oxygen and heat, respectively (see Supplement~S8 for more details). Thus, although no physiological shunting is expected for oxygen exchange, heat exchange could be sensitive to the cord configuration.

\subsection{Configurations yielding maximum and minimum exchange}\label{sec:Nmap}
\begin{figure}
    \centering
     \includegraphics[width=0.9\textwidth]{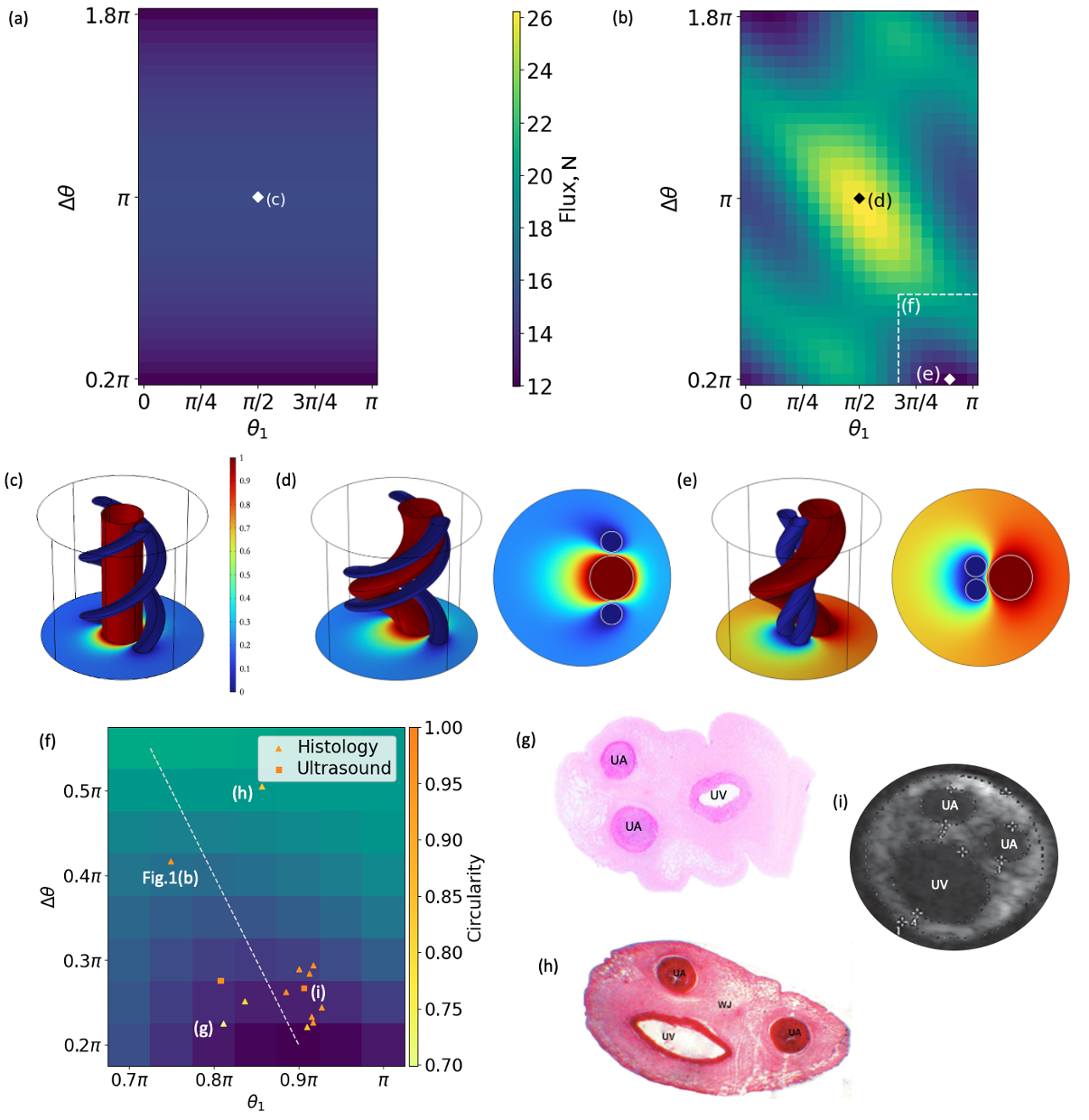}
    \caption{\textbf{Solute exchange flux $N$ for different cord configurations} in the ($\theta_1$, $\Delta\theta$) parameter space (see Fig.~\ref{fig:fetus}b). 
    The geometrical parameters used were $r_v = 0.25, r_a = 0.12, d = 0.04,\eta = 0, \Omega=3.77$ (see Table~\ref{tab:par}) for (a) $R_v = 0$ (a straight vein, as shown in (c)) and (b) $R_v = 0.21$. Cord configurations predicted by the computational model (\chg{e}) ($\theta_1 = 0.9\pi$; $\Delta \theta = 0.2\pi$) that minimise and (\chg{d}) ($\theta_1= 0.5\pi$; $\Delta \theta=\pi$) maximise solute exchange between the vessels (shown in (b)). Solute concentrations are plotted in normalised units, varying from $C=0$
    on the arterial surfaces (blue) to $C=1$
    on the surface of the vein (red). (f) Magnification of a region in (b) showing estimated angles obtained from histology and from ultrasound images of human cord cross-sections from healthy pregnancies. The line of symmetry (dashed) has a slope of $-2$ and a zero intercept at $\theta_1 = \pi$. (g) and (h) are examples of histology images with irregular shapes, circularity $<1$ (\cite{Kurakazu2019}, reproduced under CC BY 4.0, and \cite{Thomas2020}, reproduced by permission of Taylor \& Francis Ltd, tandfonline.com). 
    (i) Ultrasound image of a cord with approximately unit circularity (\cite{Kurita2009}, reproduced by permission of Karger Publishers, \copyright\,2009).}
    \label{fig:fmaps}
\end{figure}
We now consider cross-sectional cord configurations with two arteries in different positions by changing the angles $\theta_1$ and $\Delta \theta = |\theta_2 -\theta_1|$ (Fig.~\ref{fig:fetus}b). We keep the artery-vein distance ($d= 0.04$) constant and use $\theta_1 \in [0,\pi]$, while $\theta_2 \in [0,2\pi)$. We plot the exchange flux $N$ for all combinations of $\theta_1$ and $\theta_2$ for tightly coiled ($\Omega = 3.77$) configurations with a straight vein and zero offset $R_v = 0$ (Fig.~\ref{fig:fmaps}a,c) and for a coiled vein with an offset from the cord centreline $R_v = 0.25$ (Fig.~\ref{fig:fmaps}b,d). Configurations with overlapping arteries were excluded from the map; the smallest possible difference in angles such that the arteries do not intersect was $\Delta \theta = 0.2\pi$. 

When the vein is located at the centre of the cord ($R_v=0$; Fig.~\ref{fig:fmaps}c), the exchange flux is invariant with respect to $\theta_1$ and depends only on the relative position $\Delta\theta$ of the two UAs around the UV (Fig.~\ref{fig:fmaps}a). 
The exchange is maximised at $\Delta\theta=\pi$ when the arteries are aligned on opposite sides of the vein (Fig.~\ref{fig:fmaps}c).
Increasing the offset (helical radius) of the vein from the cord's centre ($R_v>0$), and hence twisting the vein, amplifies the solute flux (the maximum $N$ increases from about 15 to 26; see Fig.~\ref{fig:fmaps}b). Furthermore, the displaced vein, which acts as a source,  %
breaks the radial symmetry of the concentration field and produces a complex landscape of enhanced and reduced values for the exchange flux, as shown in Fig.~\ref{fig:fmaps}(b). The region around the global maximum ($\theta_1\approx0.5\pi$ and $\Delta\theta \approx \pi$) corresponds to the configuration where the arteries are aligned on either side of the vein, perpendicular to the $x$-axis that passes through the centre of the cord and centre of the vein (Fig.~\ref{fig:fmaps}d). 
Similarly, local maxima are attained when the arteries are close to each other in the region $\theta_1 \in [\pi/4,\pi/2]$. The global minimum ($\theta_1 \approx 0.9\pi$ and $\Delta\theta \approx 0.2\pi$) is achieved when arteries are close to each other and are placed along the $x$-axis near the centre of the cord (Fig.~\ref{fig:fmaps}e). 

The configurations that maximise the exchange flux in both scenarios (Fig.~\ref{fig:fmaps}c,\chg{d}) are associated with UAs that \chg{straddle} the UV, allowing for a larger surface area. However, a more general map in Fig.~\ref{fig:fmaps}(b) reveals that the exchange flux is smaller when the two UAs are aligned close to the $x$-axis ($\theta_1\approx \pi$\chg{;\, Fig.~\ref{fig:fmaps}e}) than when they are aligned orthogonally to it ($\theta_1\approx \pi/2$). This indicates the importance of a specific three-dimensional arrangement of the cord's vessels, which goes beyond the effective surface area of the UA-UV interface.

To determine realistic configurations of the human umbilical cord, we estimated the angles $\theta_1$ and $\Delta\theta$ from histological studies and ultrasound images of umbilical cords from normal pregnancies. Fig.~\ref{fig:fmaps}(g,h) shows examples of histological images of the human umbilical cord that highlight the irregular, non-circular shape of the cross-section. Shortly after birth, the umbilical arteries undergo constriction in order to prevent fetal blood loss. This effect is seen in the histological images, where the thick arterial walls close off the lumen. Similarly, the lumen of the umbilical vein collapses due to changes in pressure and presents an elliptical cross-section. In contrast, the ultrasound image in Fig.~\ref{fig:fmaps}(i) shows that an \emph{in vivo} cord has a more regular circular cross-section. The lumen of the vessels is also more visible and has a circular shape. This supports the model assumption of circular cross-sections for both the cord and the vessels, since it better represents \emph{in utero} conditions.

Fig.~\ref{fig:fmaps}(f) shows these measurements plotted in the context of the exchange flux map (Fig.~\ref{fig:fmaps}b). The observed vascular configurations approximately lie on the line of symmetry $\Delta\theta = -2\theta_1+2\pi$ for which the position of the arteries is symmetrical with respect to the $x$-axis, passing through the centre of the cord and the centre of the vein. The model indicates that this approximately symmetric arrangement of the umbilical arteries, with a small $\Delta\theta$, minimises the exchange flux in the 14 human cords from healthy pregnancies that we investigated. 

\subsection{`Virtual' \emph{vasa vasorum}}

\begin{figure}
    \centering
    \includegraphics[width = 0.7\textwidth]{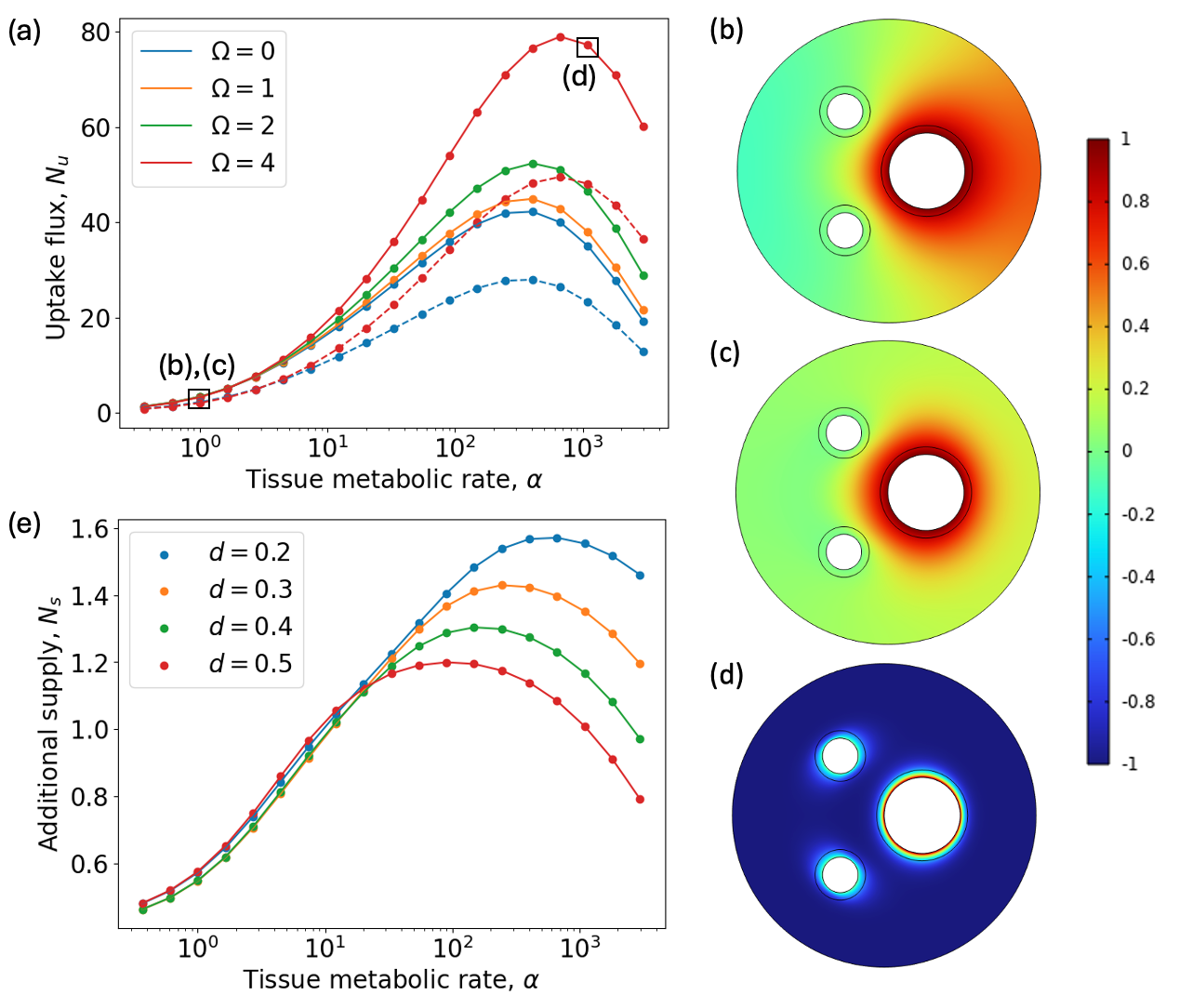}
    \caption{\textbf{The role of umbilical vascular structure in the cord tissue oxygenation.} (a) Uptake flux $N_u$ \eqref{eq:Nfluxes} by the cord tissue (excluding a thin strip of thickness $h = 0.05$ outside the vascular lumens), for varying rate of tissue metabolism $\alpha$ and cord helicity $\Omega$. The solid lines correspond to $\beta = 1$, while the dotted lines show $\beta = 2$. Scaled concentration fields for (b) $\Omega = 0;~\alpha = 1 $, (c) $\Omega = 4;~\alpha = 1$ and (d) $\Omega = 4;~\alpha = 10^3$ plotted in normalised units. (e) The contribution of the umbilical vein to the cord tissue oxygenation, quantified by the uptake flux $N_s$ relative to the flux in the case of diffusive oxygen supply by the umbilical arteries alone. Colours indicate different inter-vessel distances $d$. Other geometrical parameters used were: $R_v = 0.25$, $r_v = 0.25$, $r_a = 0.12$, $d = 0.3$, $\theta_1 = 0.8\pi$, $\theta_2 = 1.2\pi$ and $\eta = 0$.}
    \label{fig:uptake_Nrel}
\end{figure}

We hypothesise that the enhanced exchange between helical vessels might boost the supply of oxygen to the walls of the umbilical arteries and hence function as a `virtual' \emph{vasa vasorum}. In Fig.~\ref{fig:uptake_Nrel}(a) we show the effects of helicity $\Omega$, and the relative solute concentration difference between the UA and the UV $\beta$\chg{,} on the extravascular uptake flux $N_u$ (see \eqref{eq:Nfluxes}). Similarly to Fig.~\ref{fig:flux_coil}, the uptake flux increases with coiling; however, for weak tissue metabolism, the effect is less noticeable. Fig.~\ref{fig:uptake_Nrel}(b) and (c) show similar effects to Fig.~\ref{fig:flux_coil}(a), where concentration fields become more uniform as $\Omega$ increases. For higher metabolic rate $\alpha$, the impact of helicity gets stronger, and the distortions in the concentration fields in the cord tissue also have a stronger impact on the uptake flux. Fig.~\ref{fig:uptake_Nrel}(d) shows %
boundary layers around the arteries, which have an elliptical shape due to the effects of helicity on the concentration field. Additionally, the value of $\alpha$ for which the maximum flux is attained increases for larger $\Omega$.

Fig.~\ref{fig:uptake_Nrel}(a) shows a non-monotonic increase in the extravascular uptake flux $N_u$ (\ref{eq:Nfluxes}) as the metabolic rate parameter increases. Clearly, for $\alpha = 0$, the uptake flux in the cord tissue must be zero. Large $\alpha$ leads to significant spatial gradients of concentration localised in narrow regions around the vessels (Fig.~\ref{fig:uptake_Nrel}d), reducing solute exchange between the UA and the UV. We assess the impact of metabolism on the oxygen supply to the avascular cord tissue by excluding thin strips of size $h$ (comparable to the vessel wall thickness). Therefore, the uptake flux $N_u$ (Fig.~\ref{fig:uptake_Nrel}a) tends to zero for sufficiently strong metabolism ($\alpha \gg 1/h^2$), when the tissue oxygenation outside the immediate vicinity of the vascular lumen is negligible.
Fig.~\ref{fig:uptake_Nrel}(a) also illustrates the role of the relative venous--arterial concentration difference $\beta$ for the baseline case of $\beta = 1$ (oxygen concentration in the UV is twice that of the UA) and for $\beta=2$ (the UV concentration is three times that of the UA). The apparent drop in peak $N_u$ at higher $\beta$ masks an increase in the absolute uptake flux that scales with this concentration difference $C_v-C_a$ (in dimensional variables; see \eqref{eq:Nfluxes}). 

To explore the contribution of the umbilical vein to cord tissue oxygenation more directly, we consider a base case in which oxygen in the cord tissue is supplied only by the UA. We compare the net uptake flux $N_u$ to this base uptake flux $N_b$ by calculating $N_s = (N_u-N_b)/N_b$, 
which quantifies the additional oxygen supplied to the tissue due to the presence of the UV (see the Supplement \todo{S4} for more details). Fig.~\ref{fig:uptake_Nrel}(e) shows how $N_s$ varies with $\alpha$ for different vessel separation\chg{s} $d$. Similarly to Fig.~\ref{fig:uptake_Nrel}(a) there is a non-monotonic increase in excess oxygen supply.
As vessels get closer to each other, a large $\alpha$ similarly blocks the interactions between the umbilical vessels and hence diminishes the exchange between them. However, when \chg{the} vessel separation $d$ is small the additional oxygen supply to the cord tissue is almost twice as large compared to the base case (see Fig.~\ref{fig:uptake_Nrel}e).

\section{Discussion}
Beyond its obvious role as a conduit for blood enriched with oxygen and nutrients between the placenta and the fetus, rather little is known about the structure--function relationship of the umbilical cord in pregnancy. While functional observations in the cord \emph{in vivo} are typically limited to ultrasound measurements of blood flow, computational modelling~\cite{Wilke2018, Saw2017} could offer deeper insight into the physiological role of the umbilical cord. 

In this study, we examine the hypothesis that there exists a functional shunt between the umbilical vessels through a mathematical model that describes solute diffusion and uptake in the cord tissue. By exploiting a natural transformation of the coordinate system, we encapsulate the complex three-dimensional helical geometry of the umbilical vessels in a computationally efficient and interpretable two-dimensional problem, which is formulated in the cross-sectional plane of the cord.
The model reveals that cord coiling and vascular proximity promote solute exchange between the umbilical vessels (Fig.~\ref{fig:flux_coil}). In the limit of moderate helicity ($\Omega \le 1$), the model shows that the exchange flux is approximately proportional to the square of the cord helicity, or equivalently the UCI augmented by the cord diameter $D$:\, $\Omega^2 \sim D^2(\text{UCI})^2$. Similarly, in the limit of a thin gap between the umbilical vessels, separated by the relative distance $d \ll 1$, the model predicts that the dominant contribution to the exchange flux is $\mathcal{O}(d^{-1/2})$. Similar to the earlier results for blood flow modelling in the umbilical cord~\cite{Wilke2018,WilkeMattner_etal21}, these geometric determinants show that the UCI alone does not capture the essential features of solute exchange and thus can be a poor indicator of clinical pathology.   

The simulations also reveal a non-trivial relationship between the vascular configurations in the cross-section of the umbilical cord and the extent of solute exchange between the vessels (see Fig.~\ref{fig:fmaps}). We %
use theoretical predictions to interpret the structural images of cords from healthy pregnancies. In all the samples we investigated, the umbilical arteries were positioned approximately symmetrically with respect to the axis, passing through the cord centre and the centre of the vein, so that the observed configurations tend to minimise shunting between the umbilical vessels. 
To put this in the context of the typical transported solutes, we assessed the flux exchanged between the umbilical vessels relative to the flux delivered to the fetus by the umbilical vein, as quantified by a Damk\"{o}hler number $\mathrm{Da}$ (see \eqref{eq:Da}). %
Oxygen transport in the umbilical cord is diffusion-limited (advection-dominated, as represented by a small $\mathrm{Da}$), implying that the amount of oxygen exchanged between the cord vessels can be neglected. In contrast, the diffusive coupling for heat transfer may be significant, because the relevant $\mathrm{Da}$ is significantly larger. Given the tendency to minimise this coupling in the measured cord configurations, it opens an intriguing possibility that the cord structure can be partially explained by the demands of fetal thermoregulation.

Although the human umbilical cord tissue lacks \emph{vasa vasorum}~\cite{Davies2017}, 
the cord's WJ contains mesenchymal stem cells (MSCs), fibroblasts, and other connective tissue cells \cite{Benirschke2012}, all of which are metabolically active and depend on oxygen to perform their functions, which are essential for the health of the cord (such as tissue repair by the MSCs and collagen synthesis by the fibroblasts). Our model demonstrates how the functional shunt between the umbilical vessels can act as `virtual' \emph{vasa vasorum} by amplifying the oxygen supply to the cord tissue. While the cells residing in the WJ are likely adapted to a relatively low oxygen environment compared to other tissues \cite{Davies2017}, we hypothesise that abnormal coiling and vascular separation distance could adversely affect the development and function of the perivascular cord tissue; however, further experimental studies are needed to evaluate the physiological significance of this mechanism. 

Our modelling approach has many limitations and assumptions that warrant further study. For example, we neglected the effects of potential solute heterogeneity on the surface of umbilical vessels (which is important in parallel uncoiled exchanger systems \cite{Pierre2014}). 
However, the length of the cord and the helicity-induced secondary flows \cite{Dean1928,Zabielski1998} will promote the mixing of solutes in umbilical vessels, justifying to some extent the azimuthally uniform approximation. We also neglected axial concentration gradients and assumed a constant coiling pitch, whereas human cords can exhibit slow pitch variations~\cite{Wilke2018}. Future work can generalise this model by considering non-zero axial concentration gradients to study exchange along the whole cord and explore more accurately how the functional shunt affects the transport of solutes between the placenta and the fetus. 

Additionally, in this study, we primarily analysed the cord structure based on histological images; however, the \emph{in vivo} structure of the cord is prone to significant perturbations post-delivery. For example, the umbilical vein lumen collapses due to changes in pressure, while the umbilical artery lumen contracts and clots within it help avoid neonatal blood loss \cite{Nandadasa2020}. These effects can be seen in Fig.~\ref{fig:fmaps}(g,h), where the vein has an elliptical shape, the arterial lumens are not readily visible, and the cord's cross-section can have an irregular (non-circular) shape. In contrast, ultrasound images better reproduce the \emph{in utero} environment since the vessels are not collapsed or constricted, with more circular shapes and visible lumens 
(Fig.~\ref{fig:fmaps}i). Our analysis relies on the relative centre-of-mass locations of the umbilical vessels, rather than finer shape details, making our results less sensitive to these \emph{ex vivo} distortions. Nevertheless, in future studies, ultrasound images should be used for a more accurate representation of the umbilical vascular structure. 

We have also limited our analysis to cords from normal human pregnancies. Pathological studies often report thin cords with a significant decrease in WJ~\cite{Debebe2020}, and instances of cords with a single UA, which have been associated with fetal growth restriction, pre-eclampsia and other adverse pregnancy outcomes~\cite{Bruch1997}. Similarly, the morphology of the umbilical cord is known to vary significantly across different placental mammals~\cite{Benirschke2007}. These structural differences motivate the need for a systematic quantitative comparison of the cord's performance in healthy and pathological human pregnancies (using multiple imaging modalities \cite{ClarkJames21}), as well as its interspecies variations \cite{Laundon_etal24,Bappoo_etal24}.

To conclude, our study offers a new perspective on the structure--function relationship of the umbilical cord by investigating solute exchange between umbilical vessels. Using image-based mathematical modelling, we have demonstrated how cord coiling can significantly enhance diffusive coupling between vessels, while the relative vascular arrangement in human cords typically minimises this coupling. These findings have potential implications for the thermal regulation of the fetus and the oxygenation of the cord tissue. The approach outlined in this paper could be expanded to address the impact of this functional shunt on fetal development in human pregnancy pathologies and across different~species.

\vspace{1em}

\ethics{This study did not require ethical approval.}

\dataccess{All data needed to evaluate the conclusions in the paper are present in the paper and/or the Supplementary Materials. The associated structural datasets and computational codes can be accessed via the \chg{Figshare repository \url{https://doi.org/10.6084/m9.figshare.28382012}}. 
}

\aucontribute{TW, OEJ and ILC conceptualised the study and co-developed the methodology; EDJ supervised data collection and contributed to the interpretation of the results; SNS supervised image processing and contributed to the interpretation of the results; TW developed the codes, performed the simulations and image analysis, and drafted the initial manuscript. All authors discussed the results and contributed to the writing of the manuscript.}

\competing{The authors declare no competing interests.}

\funding{This work was supported by \chg{an} A*STAR Scholarship Award to TW. EDJ, OEJ \& ILC also gratefully acknowledge partial support by the Wellcome Leap \emph{In Utero} Programme.}

\ack{The authors would like to thank Prof.\ Hwee Kuan Lee and Dr Nicholas Cheng (A*STAR Bioinformatics Institute) for helpful discussions.}

\bibliography{references.bib}
\bibliographystyle{RS}

\end{document}


\title{\Large{A functional exchange shunt in the umbilical cord: \\ the role of coiling in solute and heat transfer}\\[0.5em]
\Large{SUPPLEMENTARY MATERIAL}\\[0.5em]}

\author{Tianran~Wan}
\affiliation{Department of Mathematics, University of Manchester, Manchester M13 9PL, UK}
\affiliation{Bioinformatics Institute, A*STAR, 138671, Singapore}
\author{Edward~D.~Johnstone}
\affiliation{Maternal and Fetal Health Research Centre, University of Manchester, Manchester, M13 9WL, UK}
\author{Shier~Nee~Saw}
\affiliation{Department of Artificial Intelligence, Universiti Malaya, Kuala Lumpur, 50603, Malaysia}
\author{Oliver~E.~Jensen}
\affiliation{Department of Mathematics, University of Manchester, Manchester M13 9PL, UK}
\author{Igor~L.~Chernyavsky}
\affiliation{Department of Mathematics, University of Manchester, Manchester M13 9PL, UK}
\affiliation{Maternal and Fetal Health Research Centre, University of Manchester, Manchester, M13 9WL, UK}

\maketitle
\onecolumngrid

\section{A three-dimensional model of solute exchange between helical vessels} \sectionmark{3D model}

\begin{figure}
    \centering
    \includegraphics[width=0.9\textwidth]{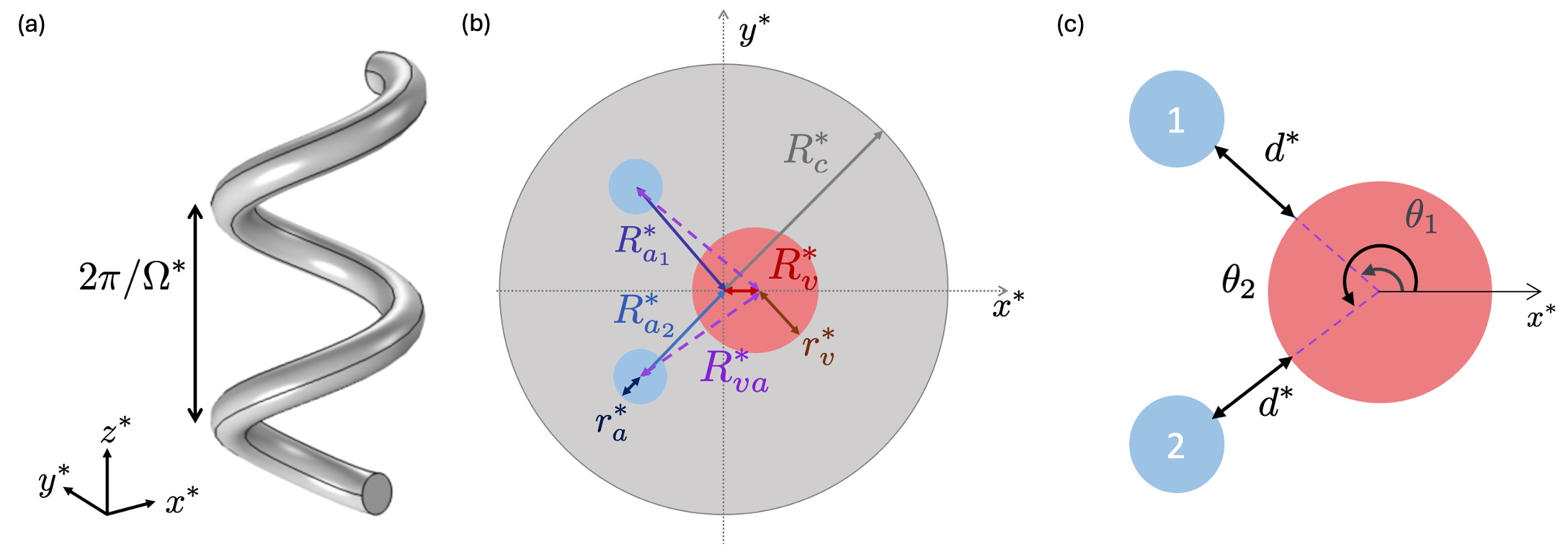}
    \caption{Diagram showing (a) the pitch $2\pi/\Omega^*$ of an internal vein or artery. (b) Cross-section of the cord at $\chg{\zeta}=0$ with radius $R^*_c$ showing the position of the vein (red) and the arteries (blue). $R^*_v$ and $R^*_{a_i}$ ($i=1, 2$) are the helical radii of the vein and arteries respectively and $r^*_v$ and $r^*_a$ are the vessel radii of the vein and arteries respectively. The distance between the centre of the vein and the centre of each artery is $R^*_{va}$. (c) Diagram of the cross-section of the cord in the plane $z^*=0$ \chg{($\zeta=0$)} showing the position of the arteries determined by $\theta_i$ for $i=1,2$, the angle of artery $i$ with respect to the $x^*$ axis on the plane $z^*=0$, and the separation distance $d^*$ between the vein and arteries. \chgrev{The cross-section in (b) is rotated about $x^* = 0$, $y^* = 0$ as $z^*$ increases, in order to define the 3D vessel geometries.}}
\label{fig:geom_cord}
\end{figure}

We build a three-dimensional model of solute exchange between helical vessels in an umbilical cord to quantify the amount of solute exchanged. We solve the diffusion equation in the umbilical cord accounting for its helical structure and calculate the flux of solute between vessels. 

We model the umbilical cord as a straight circular cylinder of radius $R^*_c$\chg{, which} contains three helical vessels (one \chg{umbilical} vein\chg{, UV,} and two \chg{umbilical} arteries\chg{, UA}). \chg{We further assume that} each \chg{vessel has} a circular cross-section in the plane perpendicular to the axis of the cylinder and a constant pitch. Fig.~\ref{fig:geom_cord}(a) shows the pitch $2\pi/\Omega^*$, defined as the height of one complete helix turn measured parallel to the axis of the cylinder. Let $\mathbf{x}^*=(x^*,y^*,z^*)$ be a dimensional Cartesian coordinate system such that $\mathbf{\hat{z}}$ lies along the axis of the cylinder.
The vessels are built by sweeping the centre of a circle along a helix defined for the \chg{umbilical} vein \chg{($v$)} and the arteries (\chg{$a$}, with $i=1,2$) as
\begin{subequations}
\begin{equation}
  \chg{x^*_v(\zeta) = R^*_v\cos{\zeta}\,, \quad y^*_v(\zeta) = R^*_v\sin{\zeta}\,, \quad z^*_v(\zeta) = \zeta/\Omega^*\,,} 
\end{equation}
\begin{equation}
    \chg{x^*_{a_i}(\zeta) = R^*_{va}\cos(\zeta+\theta_i) + R^*_v\cos{\zeta}\,,\quad y^*_{a_i}(\zeta) = R^*_{va}\sin(\zeta+\theta_i)+R^*_v\sin{\zeta}\,, \quad  z^*_{a_i}(\zeta) = \zeta/\Omega^*\,,}
\end{equation}
\end{subequations}
for $\chg{\zeta} \in [0, \Omega^*L^*]$. Here $\chg{\zeta} =0$ is the fetal end of the cord and $\chg{\zeta} = \Omega^*L^*$ is the maternal end of the cord, with $L^*$ being the length of the cord. The helical radius, the distance between the centre of the cord and the centre of the vessel, is $ R^*_v$ or $R^*_{a_1}$, $R^*_{a_2}$ for each vessel respectively (see Fig.~\ref{fig:geom_cord}b). $\theta_i$ for $i=1,2$ \chgrev{is} the angle subtended by the centre of the artery $i$ with the $x^*$ axis that passes through the centre of the cord and the centre of the vein (see Fig.~\ref{fig:geom_cord}c). Additionally, we define the angle between the arteries as $\Delta\theta = |\theta_2-\theta_1|$. We define the distance between the centre of the vein and the centre of each artery to be $R^*_{va} =d^*+r^*_v+r^*_a$\chg{,} where $d^*$ is the minimum distance between each artery and the vein; $r^*_v$ and $r^*_a$ are the radii of the vein and artery cross-sections respectively. We assume that $R^*_{va}$ and $r^*_a$ are the same for both arteries. Fig.~\ref{fig:geom_cord}(b) \chg{shows t}he position of the vessels in the cross-section of the cord in the plane $z^*=0$ \chg{($\zeta=0$)}. The other \chg{planar} cross-sections \chg{$z^*= \text{const}$} have the same \chg{vascular configuration} rotated \chg{by $\zeta$}. The centre of the cord lies along $(x^*,y^*,z^*) = (0,0,z^*)$. The centre\chg{lines} of the vein \chg{and} the arteries \chg{($i=1,2$) pass through} $(R^*_v, 0,0)$ and $(R^*_{va} \cos(\theta_i)+R^*_v, R^*_{va} \sin(\theta_i),0)$\chg{, respectively, at $\zeta=0$,} where $R^*_{a_i}= \sqrt{(R^*_{va})^2+2R^*_v R^*_{va}\cos(\theta_{i})+(R^*_v)^2}$ \chg{(Fig.~\ref{fig:geom_cord}b)} 
\chgrev{so that $x^*_{a_i}(\zeta) = R^*_{a_i}\cos(\zeta+\psi_i)$,\,  $y^*_{a_i}(\zeta) = R^*_{a_i}\sin(\zeta+\psi_i)$, where $\psi_i$ is the angle subtended with the $x^*$--axis at $x^*=0,\, y^*=0$}. \chg{Table~1 of the main text} \chg{gives the} values and corresponding source\chg{s} for these geometrical parameters. 

\begin{figure}
    \centering
    \includegraphics[width=0.9\textwidth]{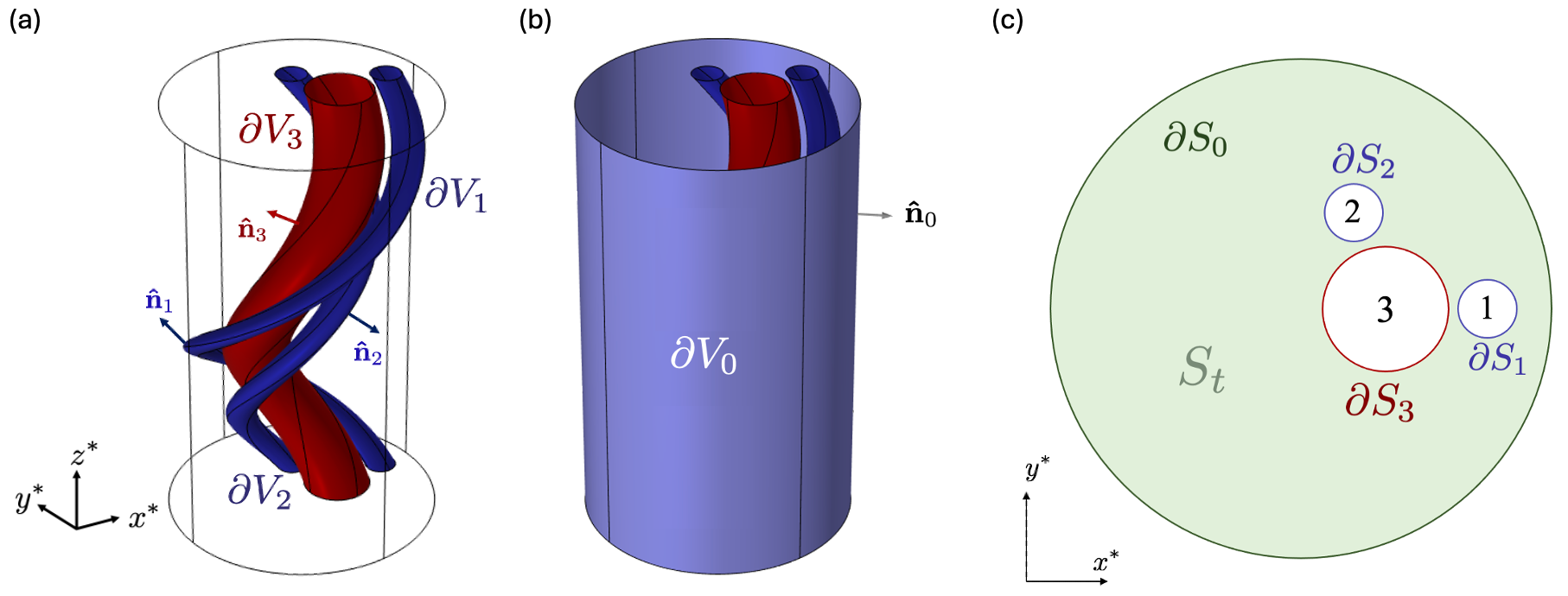}
\caption{Diagram showing (a) $\mathbf{\hat{n}}_i$, the unit normal out of vessels' boundary $\partial V_i$ for $i = 1,2,3$. (b)  $\mathbf{\hat{n}}_0$, the unit normal out of $\partial V_0$ the 3D surface  (c) $\partial S_i$ the vessel cross-sectional boundary in the $x^*,y^*$ plane for fixed $z^*$, and $S_t$ the cord tissue.} %
\label{fig:3Dgeom_cord}
\end{figure}

Let $V_i$ for $i=0,1,2,3$ be the domain occupied by the cord, the two arteries and the vein respectively, and define $\partial V_i$ to be the corresponding boundaries as shown in Fig.~\ref{fig:3Dgeom_cord}(a). Then the cord tissue volume can be defined as $V_t \equiv V_0 \setminus (V_1\cup V_2\cup V_3)$. Denote a plane $\mathcal{P}$ where $z$ is constant. Then the line boundaries for the cord and each vessel on the $(x^*,y^*)$ plane for fixed $z^*$ can be written as $\partial S_i \equiv \partial V_i \cap \mathcal{P}$ for $i = 0,1,2,3$ respectively. The cord tissue surface on the $(x^*,y^*)$-plane for fixed $z^*$ is defined by $S_t = V_t \cap \mathcal{P} = S_0\setminus(S_1 \cup S_2 \cup S_3)$, and we denote the end plates of the cord by $\chg{S_f}\equiv V_t \cap \mathcal{P}\big|_{z^*=0} $ and $\chg{S_p} \equiv V_t \cap \mathcal{P}\big|_{z^*=L^*}$ respectively (see Fig.~\ref{fig:3Dgeom_cord}c).

The solute (or heat) concentration, $c^*(\mathbf{x}^*)$, in the cord tissue $V_t$, is assumed to obey the steady diffusion\chg{--uptake} equation
\begin{equation}\label{eq:uptake_dim}
    D_t^*\nabla^2 c^* = q^*_0 c^* \quad \text{in $V_t$}.
\end{equation}
Here $q^*_0$ is the rate of solute uptake by the cord's tissue, which is assumed to be uniform. For oxygen metabolism \chg{at relatively low concentrations in the cord tissue}, we assume that $q^*_0 \approx q^*_\mathrm{max}/c^*_{50}$, where $q^*_\mathrm{max}$ is the maximum rate of oxygen metabolism and $c^*_{50}$ is the concentration at which the metabolic rate reaches 50\% of its maximum (see \chg{\cite{Erlich19-metabolism} and Table~\ref{tab4:D} below} for more details).

Since the umbilical vessels are helical, we need to consider the effects of curvature on internal fluid flow. The most evident consequence of curvature is the presence of secondary flow in the form of vortices caused by centrifugal forces \citep{Dean1928, Zabielski1998}. 
We will assume that these secondary flows promote the mixing of the vessels' concentration, such that we can assume uniform concentration around each vessel within the cross-sectional plane. Furthermore, we assume zero axial concentration gradient, providing the boundary conditions
\begin{equation}
    c^* = c^*_i \text{ on } \partial V_i \quad \text{for $i$ = 1, 2, 3}, 
\end{equation}
where $c^*_1$, $c^*_2$ and $c^*_3$ are the fixed concentrations in the two arteries and the vein respectively.

On the outer cord surface, $\partial V_0$, excluding the end plates, we impose a mixed boundary condition 
\begin{equation}\label{eq:dimbc0}
    \chg{\eta^*}(c^*- \chg{c^*_{0}}) + (\mathbf{\hat{n}}_0 \cdot \nabla c^*)  = 0 \quad \text{on} \quad \partial V_0 %
\end{equation}
for some constant $\eta^*$ and fixed external concentration $c^*_0$. 
Here, $\mathbf{\hat{n}}_0$ is the unit normal oriented out of the outer surface of the cord $\partial V_0$ (see Fig.~\ref{fig:3Dgeom_cord}b). When \chg{$\eta^*$ is small}, \eqref{eq:dimbc0} specifies a flux on the outer boundary, while \chg{large $\eta^*$} specifies a fixed concentration.
\chg{We further assume $c^*_1=c^*_2 \equiv c^*_a$,\, $c^*_3 \equiv c^*_v$,\, $c^*_0 \equiv c^*_a$, neglecting possible concentration differences between the two umbilical arteries, or between the arterial blood and the amniotic fluid.}

At the end plates $\chg{S_f}$ and $\chg{S_p}$, we impose no diffusive flux 
\begin{equation}
    \mathbf{\hat{n}} \cdot \nabla c^* = 0 \quad \text{on $\chg{S_f}$ and on $\chg{S_p}$.} %
\end{equation}
The total flux entering the cord across $\partial V_0$ and entering each internal vessel is defined as
\begin{equation}
    J^*_i = D_t^*\int_{\partial V_i} \mathbf{\hat{n}}_i\cdot \nabla c^*  \mathop{}\!\mathrm{d}A\,,\; \text{where\; $i = 0,1,2,3$,} %
\end{equation}
where $D_t^*$ is the diffusion coefficient and $\mathbf{\hat{n}}_i$ is the unit outward normal vector to surface $\partial V_i$.

\subsection*{Non-dimensionalisation}
We scale all the geometrical parameters using $R^*_c$ as a reference length and obtain dimensionless geometric parameters (without asterisks) in Table 1. 
\chg{W}e scale all the concentrations such that 
\begin{equation}\label{eq:scaling}
        c^*(x^*,y^*,z^*) =c^*_a + (c^*_v -c^*_a)\, c(x,y,z)
\end{equation}
and write
\begin{equation}
    \chg{\Omega^*=\Omega/R^*_c,\,} \quad \eta^* = \eta\, R^*_c, \quad J^*_i = D_t^* R^*_c \,(c^*_v-c^*_a)\, J_i,\quad \alpha = \frac{q^*_0\, (R^*_c)^2}{D_t^*} \quad \text{ and } \quad \beta =\frac{c^*_v-c^*_a}{c^*_a}.
\end{equation}
The governing equations become
\begin{subequations}\label{eq4:3Dc}
\begin{align}
        \nabla^2 c &= \alpha \left(c +\frac{1}{\beta}\right) \quad \text{in $V_t$}, \label{eq4:3difcart}\\
    \chg{\eta}\,c +  (\mathbf{\hat{n}}_0 \cdot \nabla c)  &= 0 \quad \text{on } \partial V_0 ,\label{eq4:outbc}\\ %
    \mathbf{\hat{n}} \cdot \nabla c&= 0 \quad \text{on \chg{$S_f$, $S_p$}},\\
    c(x,y,z) &= 1 \quad\text{on } \partial V_3 \;\text{ and }\; c(x,y,z) = 0 \text{ on } \partial V_i
    \text{ for $i$ = 1, 2}. \label{eq4:3DCi}
\end{align}
\end{subequations}
We first consider cases when tissue metabolism is absent ($\alpha=0$) then in section \ref{app:uptake} we analyse the effects of tissue metabolism for $\alpha>0$.

The total flux entering the cord across $\partial V_0$ and entering each internal vessel is defined as 
\begin{equation}\label{eq4:Ni}
    J_i = \int_{\partial V_i} \mathbf{\hat{n}}_i\cdot \nabla c  \mathop{}\!\mathrm{d}A\,,\; \text{where\; $i = 0, 1,2,3$.} %
\end{equation}
Here, the outward unit normal vector to the surface defined as $f(x,y,z) = 0$ is given by $\mathbf{\hat{n}} = \nabla f / |\nabla f|$. On the cord boundary $\partial V_0$, $f(x,y,z)=x^2+y^2-1$ so that $\mathbf{\hat{n}}_0 = (x,y,0)$. The other normal vectors, $\mathbf{\hat{n}}_i$ for $i = 1,2,3$, typically have a component that lies outside any plane $z=$ constant (see Fig.~\ref{fig:3Dgeom_cord}a).  

\section{Exchange problem in helical coordinates}\label{4suppl:hel}
We now introduce a new coordinate system $\mathbf{X}=(X,Y,Z)$ such that the vessels' (circular) cross-sections in any plane $\mathcal{P}$ lying perpendicular to the axis of the cord remain fixed with respect to $(X,Y,Z)$. This requires the new frame to rotate at a rate $\Omega$ as one moves along the cord. We define the helical coordinates as
\begin{equation}\label{hel_coord}
    X = x\cos{\left(\Omega z\right)} + y \sin{\left(\Omega z\right)}, \quad
     Y = -x\sin{\left(\Omega z\right)} + y\cos{\left(\Omega z\right)}, \quad
     Z = z .
\end{equation}
Expressing the concentration as $c(x,y,z) = C(X,Y,Z)$ \chg{and denoting $C_X\equiv \partial_XC$}, we can \chg{express} the derivatives
\begin{subequations}
    \begin{align}
            c_x &= C_X \cos{\left(\Omega Z\right)} - C_Y \sin{\left(\Omega Z\right)},\\
    c_y &= C_X \sin{\left(\Omega Z\right)} \chg{\,+\,} C_Y \cos{\left(\Omega Z\right)}, \\
    c_z &= \Omega Y C_X -\Omega X C_Y + C_Z, 
    \end{align}
\text{so that}
\begin{equation}\label{eq:hel_chain}
    \nabla \chg{\equiv \mathbf{\hat{x}}\partial_x+\mathbf{\hat{y}}\partial_y+\mathbf{\hat{z}}\partial_z} \equiv \mathbf{\hat{x}}[\cos(\Omega Z)\partial_X-\sin(\Omega Z)\partial_Y]+\mathbf{\hat{y}}[\sin(\Omega Z)\partial_X+\cos(\Omega Z)\partial_Y]+\mathbf{\hat{z}}[\partial_Z+\Omega(Y\partial_X-X\partial_Y)].
\end{equation}
\end{subequations}
We can reduce the 3D problem \eqref{eq4:3Dc} to 2D by expressing it in helical coordinates and seeking a solution that is independent of Z satisfying
\begin{align}\label{eq:2Dc}
    \nabla^2 c &= c_{xx} + c_{yy} \chg{\,+\, c_{zz}} \\ \nonumber
    &= C_{XX} + C_{YY} +\Omega^2(Y^2C_{XX}+X^2C_{YY}-XC_X-YC_Y-2XYC_{XY})\\ \nonumber
    &\equiv \nabla_{\perp} \circ \left[\nabla_{\perp} C + \Omega^2 \mathbf{X}^\perp(\mathbf{X}^\perp\circ \nabla_\perp C)\right] = \alpha(\chg{C}+1/\beta), 
\end{align} 
where $\nabla_\perp \equiv (\partial_X,\partial_Y,0)$, $\mathbf{X}^\perp \equiv (Y,-X,0)$ and the product $\circ$ is defined as $(a,b,c) \circ (d,e,f) \equiv ad + be +cf$. In general, this is not equivalent to a scalar product since the helical coordinate system $(X,Y,Z)$ is not orthogonal. The term proportional to $\Omega^2$ shows how in-plane azimuthal concentration gradients drive azimuthal fluxes that are amplified at increasing radial distances from the centre of the cord.

Consider a surface defined by the function $F(\mathbf{X}_{\perp}) = 0$, where $\mathbf{X}_{\perp} \equiv (X,Y\chg{,0})$. The normal $\chg{\mathbf{n} =\,} \nabla F$ is obtained using \eqref{eq:hel_chain} as
    \begin{align}\label{eq:n}
    \mathbf{n} &= \mathbf{\hat{x}}[F_X \cos(\Omega Z)-F_Y\sin(\Omega Z)]+\mathbf{\hat{y}}[F_X \sin(\Omega Z)+F_Y\cos(\Omega Z)]+\mathbf{\hat{z}}\Omega[YF_X -XF_Y].
\end{align}
We normalise this by calculating 
\begin{equation}
    |\mathbf{n}| = \sqrt{|\nabla_\perp F|^2 + \Omega^2 (YF_X -XF_Y)^2} \equiv \sqrt{|\nabla_\perp F|^2 + \Omega^2 (\mathbf{X}^\perp \circ \nabla_\perp F)^2}.
\end{equation}
To derive a concentration gradient normal to $F=0$ in terms of helical coordinates, we combine \eqref{eq:n} with \eqref{eq:hel_chain} to give
\begin{equation}
    \mathbf{\hat{n}} \cdot \nabla = \frac{F_X\partial_X+F_Y\partial_Y +\Omega^2[Y^2F_X\partial_X+X^2F_Y\partial_Y-XY(F_X\partial_Y+F_Y\partial_X)]\chg{\,+\, \Omega(YF_X-XF_Y)\partial_Z}}{|\mathbf{n}|},  
\end{equation}
\chg{which reduces to}
\begin{equation}\label{eq:3Dflux}
    \mathbf{\hat{n}} \cdot \nabla C = \frac{\chg(\nabla_\perp F\chg{)}\circ [\nabla_\perp C+\Omega^2\mathbf{X}^\perp(\mathbf{X}^\perp \circ \nabla_\perp C)]}{|\mathbf{n}|}\,,
\end{equation}
\chg{as we are looking for a solution independent of $Z$, with $\partial_Z C = 0$.}
On the outer boundary $\partial S_0$, where $F_0(\mathbf{X}_{\perp}) = X^2+Y^2-1=0$, we have $\nabla_\perp F_0 \circ \mathbf{X^\perp}= \mathbf{X}_{\perp}\circ\mathbf{X^\perp}=0$\chg{, and} \eqref{eq:3Dflux} gives 
\begin{equation}
     \mathbf{\hat{n}}_0 \cdot \nabla C = \frac{\nabla_\perp F_0}{|\nabla_\perp F_0|}\circ \nabla_\perp C =  \mathbf{X}_{\perp} \circ \nabla_\perp C.
\end{equation}
Then the condition \eqref{eq4:outbc} across the outer boundary of the cord becomes
\begin{equation}\label{eq:robin}
    \chg{\eta}\,C + (\mathbf{X}_\perp \circ \nabla_\perp C) = 0 \quad \text{on $\partial V_0$}.
\end{equation}

In order to obtain an expression for the diffusive flux across $\partial S_n$, $n=1,2,3$ in the new coordinate system, we need to transform the area element $\d A$ in \eqref{eq4:Ni}.  We parametrise a circle with centre $(X_0,Y_0)$ and radius $r$ in the $(X,Y)$ plane by
\begin{equation}\label{eq:circle}
    X = X_0 + r \cos{\phi}, \quad Y = Y_0 + r \sin{\phi} \quad \text{for $0\leq\phi<2\pi$},
\end{equation}
so that 
\begin{equation}\label{eq:circle_parder}
    X_\phi = -r \sin{\phi} = -(Y-Y_0) \quad \text{and} \quad Y_\phi = r \cos{\phi} = X-X_0.
\end{equation}
The coordinate transformation \eqref{hel_coord}, can be written as 
\begin{equation}
    \mathbf{x}= \begin{pmatrix}
x \\
y \\
z 
\end{pmatrix} =
    \begin{pmatrix}
\cos(\Omega Z) & -\sin(\Omega Z) & 0 \\
\sin(\Omega Z) & \cos(\Omega Z) & 0 \\
0 & 0 & 1 
\end{pmatrix}  \begin{pmatrix}
X \\
Y \\
Z 
\end{pmatrix}. 
\end{equation}
The surface $S$ of the helical vessel having \eqref{eq:circle} as its cross-section in a plane $Z=$ constant is parametrised by $\phi$ and $Z$ so that 
\begin{equation}\label{eq:x_S}
    \mathbf{x}|_S =     \begin{pmatrix}
\cos(\Omega Z) & -\sin(\Omega Z) & 0 \\
\sin(\Omega Z) & \cos(\Omega Z) & 0 \\
0 & 0 & 1 
\end{pmatrix}  \begin{pmatrix}
X_0 + r \cos{\phi} \\
Y_0 + r \sin{\phi} \\
Z 
\end{pmatrix}. 
\end{equation}
We seek the vectors $\partial\mathbf{x}|_S/\partial \phi$ and 
$\partial\mathbf{x}|_S/\partial Z$. First, using \eqref{eq:circle_parder}, 
\begin{align}
    \frac{\partial\mathbf{x}|_S}{\partial \phi} &=     
\begin{pmatrix}
\cos(\Omega Z) & -\sin(\Omega Z) & 0 \\
\sin(\Omega Z) & \cos(\Omega Z) & 0 \\
0 & 0 & 1 
\end{pmatrix}\begin{pmatrix}
 X_\phi\\
Y_\phi \\
0 
\end{pmatrix} %
= \begin{pmatrix}
 -\cos(\Omega Z)(Y-Y_0)-\sin(\Omega Z)(X-X_0) \\
\cos(\Omega Z)(X-X_0)-\sin(\Omega Z)(Y-Y_0)\\
0 
\end{pmatrix}.
\end{align}
Applying the product rule to \eqref{eq:x_S}, we obtain 
\begin{align}
    \frac{\partial\mathbf{x}|_S}{\partial Z} &
= \begin{pmatrix}
\cos(\Omega Z) & -\sin(\Omega Z) & 0 \\
\sin(\Omega Z) & \cos(\Omega Z) & 0 \\
0 & 0 & 1 
\end{pmatrix}
\begin{pmatrix}
 0\\
0 \\
1 
\end{pmatrix} + \begin{pmatrix}
-\sin(\Omega Z) & -\cos(\Omega Z) & 0 \\
\cos(\Omega Z) & -\sin(\Omega Z) & 0 \\
0 & 0 & 0 
\end{pmatrix} \begin{pmatrix}
X_0 + r \cos{\phi} \\
Y_0 + r \sin{\phi} \\
Z 
\end{pmatrix} \\ \nonumber
&= \begin{pmatrix}
 -\chg{\Omega}(X\sin(\Omega Z)+Y\cos(\Omega Z))\\
\chg{\Omega}(X\cos(\Omega Z)-Y\sin(\Omega Z)) \\
1 
\end{pmatrix}.
\end{align}
Then
\begin{equation}\label{eq:dA-end}
    \frac{\partial \mathbf{x}|_S}{\partial \phi} \times \frac{\partial \mathbf{x}|_S}{\partial Z} = \begin{pmatrix}
    \cos(\Omega Z)(X-X_0)-\sin(\Omega Z)(Y-Y_0)\\
 \cos(\Omega Z)(Y-Y_0)+\sin(\Omega Z)(X-X_0) \\
\chg{\Omega}(XY_0 -YX_0) 
\end{pmatrix}\,.
\end{equation}
Thus we can evaluate the area element as
\begin{equation}
    \d A = \Bigg| \frac{\partial (x,y,z)}{\partial \phi} \times \frac{\partial (x,y,z)}{\partial Z}\Bigg| \d \phi \d Z 
    = \sqrt{1 + \frac{\Omega^2}{r^2}(XY_0 -YX_0)^2} \:\,r \d \phi \d Z\,.
    \label{eq:da}
\end{equation}

On the circle \eqref{eq:circle} we have
\begin{equation}
    \d X = -r\sin{\phi} \d\phi \quad \text{and} \quad  \d Y = -r\cos{\phi} \d\phi,
\end{equation}
so the arclength of the parameterised circle can be written as
\begin{equation}\label{eq:arcl}
    \d s = \sqrt{\d X^2 +\d Y^2}\, = r\d\phi. 
\end{equation}
For the surface defined by $0=F(X,Y) = \frac{1}{2} ((X-X_0)^2 +(Y-Y_0)^2-r^2)$, the normal \eqref{eq:n} becomes
\begin{multline}\label{eq:xn}
        \mathbf{n} = \mathbf{\hat{x}}[ (X-X_0)\cos(\Omega Z)-(Y-Y_0)\sin(\Omega Z)]+\mathbf{\hat{y}}[(X-X_0) \sin(\Omega Z)+(Y-Y_0)\cos(\Omega Z)]\\+\mathbf{\hat{z}}\Omega[Y(X-X_0) -X(Y-Y_0)],
\end{multline}
then the area element becomes
\begin{equation}\label{eq:area_element}
    \d A = \frac{|\mathbf{n}|}{r} \d s \d Z\,, \quad \text{where} \quad  |\mathbf{n}| = \sqrt{r^2 + \Omega^2(XY_0 -YX_0)^2},
\end{equation}
consistent with (\ref{eq:da}).

\begin{figure}
    \centering
    \includegraphics[width=0.38\textwidth]{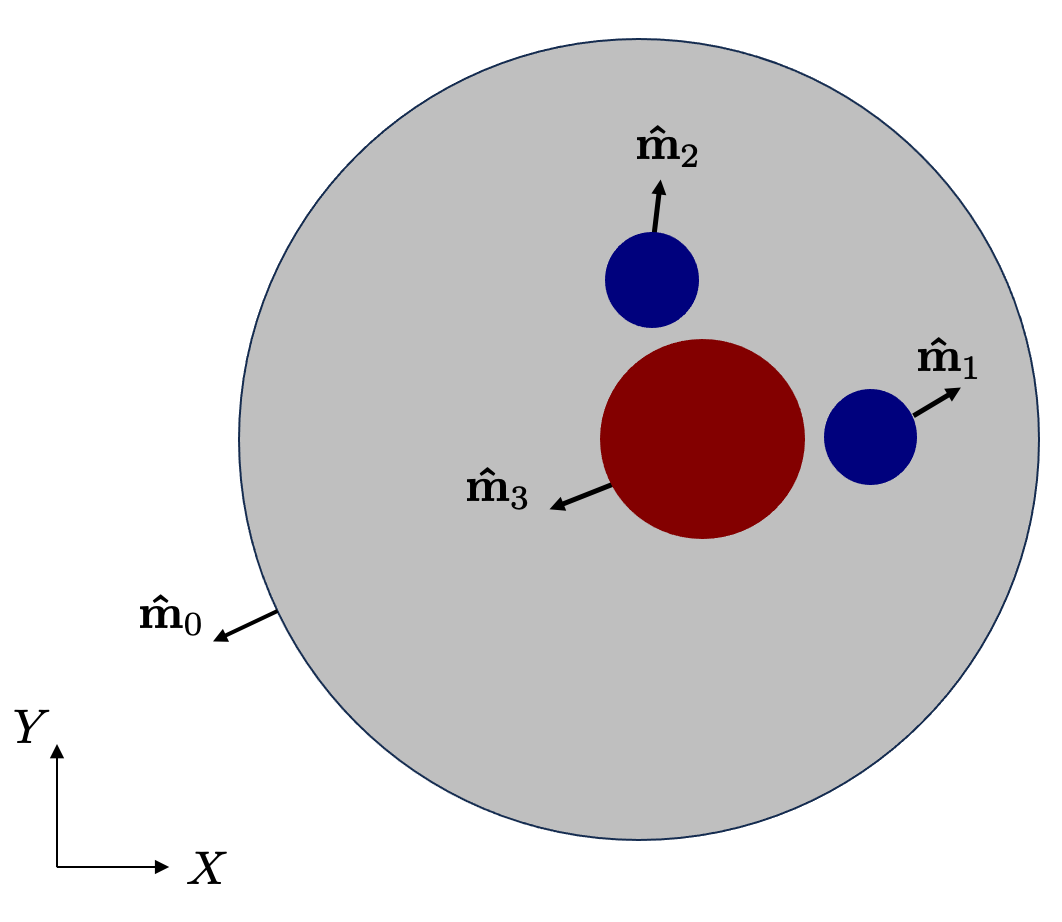}
    \caption{Diagram showing $\mathbf{\hat{m}}_i$, the in-plane unit normal out of the cord and vessels' boundary $\partial S_i$ for $i = 0,1,2,3$.}
    \label{fig4:2d_m}
\end{figure}

To calculate the flux \chg{\eqref{eq4:Ni}}, we take the line integral over the cord and the vessels' boundaries $\partial S_i$. Since the integrand is independent of $Z$\chg{, and utilising \eqref{eq:3Dflux} and \eqref{eq:area_element},} the flux can be expressed as
\begin{align}\label{eq:2dflux}
    J_i  &= \int_{0}^{\chg{L}} \int_{\partial S_i} \mathbf{\hat{m}}_i \circ \left[\nabla_{\perp} C + \Omega^2 \mathbf{H}\right]\mathrm{d}s \, \mathrm{d}Z %
    = \chg{L} \int_{\partial S_i} \mathbf{\hat{m}}_i \circ \left[\nabla_{\perp} C + \Omega^2 \mathbf{H}\right]\mathrm{d}s\\ 
    &=\chg{L}N_i \quad \text{where} \quad N_i=\int_{\partial S_i} \mathbf{\hat{m}}_i \circ \left[\nabla_{\perp} C + \Omega^2 \mathbf{H}\right]\mathrm{d}s  \quad \text{for $i=0,1,2,3$}\,, \nonumber
\end{align}
where $\mathbf{H} = (Y\,C_X - X\,C_Y)\,(Y,\,-X,\,0)$. %
The in-plane unit normals oriented out of boundaries $\partial S_i$ defined by \hbox{$F_i(\mathbf{X}_\perp)=0$} (see Fig.~\ref{fig4:2d_m}) are
\begin{equation}\label{eq:mi}
    \mathbf{\hat{m}}_i = \frac{\nabla_\perp F_i}{|\nabla_\perp F_i|}=\frac{\nabla_\perp F_i}{r_i} = \frac{1}{r_i}(X-X_{0,i},\,Y-Y_{0,i},\,0)^\mathsf{T}\,,%
\end{equation}
where $r_i\chg{\:=|\nabla_\perp F_i|}$ is the vessel radius. %
The explicit expressions of the normals for each vessel can be derived as follows. Starting with the vein, defined by 
\begin{equation}
    0=F(X,Y) = \tfrac{1}{2} \left[(X-R_v)^2 +Y^2-r^2_v)\right],
\end{equation}
using \eqref{eq:mi}, we obtain
\begin{equation}
    \hat{\mathbf{m}}_3 = \frac{\nabla_\perp F}{r_v} = \frac{1}{r_v} \begin{pmatrix}
X-R_v \\
Y \\
0 
\end{pmatrix}.
\end{equation}
For the arteries, we can define the surface as
\begin{equation}
    0=F(X,Y) = \tfrac{1}{2} \left[(X - R_{va}\cos(\theta_i) \chg{\,-\,} R_v)^2 +(Y-R_{va} \sin(\theta_i))^2-r^2_a) \right],
\end{equation}
where $i =1,2$, so the normals are
\begin{equation}
    \hat{\mathbf{m}}_i = \frac{\nabla_\perp F}{r_a} = \frac{1}{r_a} 
    \begin{pmatrix}
X-(R_{va} \cos(\theta_i)+R_v) \\
Y-R_{va} \sin(\theta_i) \\
0 
\end{pmatrix}.
\end{equation}

Consider the following identity
\begin{subequations}\label{eq:ident}
    \begin{align}
    \nabla_\perp \chg{\;\circ\;} [C \nabla_\perp C+\Omega^2C\mathbf{H}] &=\chg{ |\nabla_\perp C|^2 +C\nabla_\perp \circ [\nabla_\perp C+ \Omega^2\mathbf{H}] +\Omega^2(YC_X-XC_Y)^2} \label{eq:ident_a}\\  
    &=|\nabla_\perp C|^2 +\Omega^2(YC_X-XC_Y)^2\chg{.}
\end{align}
\end{subequations}
\chg{The second term on the \chg{right-}hand side of \eqref{eq:ident_a} becomes zero by exploiting \eqref{eq:2Dc} \chg{for} $\alpha=0$}. Then, integrating \eqref{eq:ident} over the cord's tissue in the cross-sectional plane $S_t$ we obtain
\begin{equation}\label{eq:flux_st1}
     \sum_{i=0}^3\int_{\partial S_i} \hat{\mathbf{m}} \circ [C \nabla_\perp C+\Omega^2C\mathbf{H}]\mathrm{d}s = \int_{S_t} \left\{  |\nabla_\perp C|^2 +\Omega^2(YC_X-XC_Y)^2 \right\}\mathrm{d}X\mathrm{d}Y.
\end{equation}
Assuming $\eta \,\chg{=0}$, the only contribution to the left-hand side of \eqref{eq:flux_st1} comes from the vein (for which $C=1$ on $\partial S_3$), meaning
\begin{equation}\label{eq:fluxSt}
    \int_{\partial S_3} \hat{\mathbf{m}} \circ [ \nabla_\perp C+\Omega^2\mathbf{H}]\mathrm{d}s = \int_{S_t} \left\{ |\nabla_\perp C|^2 +\Omega^2(YC_X-XC_Y)^2 \right\}\,\mathrm{d}X\mathrm{d}Y.
\end{equation}
\chg{In Section~\ref{sec:dsmall} below, we exploit the relationship \eqref{eq:fluxSt} to estimate the exchange flux in the limit when the distance between the arteries and the vein is small.}
\section{Small vessel proximity limit}\label{sec:dsmall}
\begin{figure}
    \centering
    \includegraphics[width=0.75\textwidth]{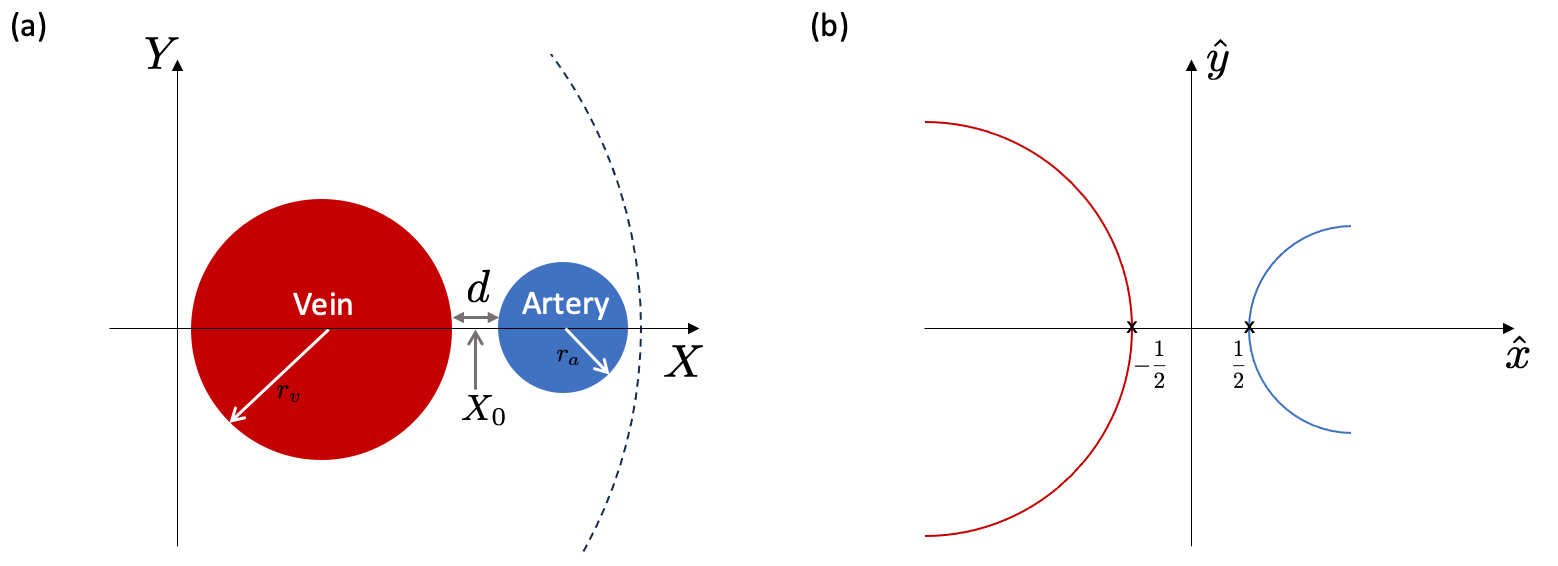}
    \caption{(a) Diagram of the vein and artery in helical coordinates highlighting the separation distance $d$ and the mid-point $X_0$. (b) Diagram of the local coordinates $(\hat{x},\hat{y})$.}
    \label{fig:smalld_coor}
\end{figure}

In this section, we explore the limit when the \chg{minimal separation} distance \chg{$d$} between the arteries and the vein is small. Define the mid-point along the line connecting the centre of the vein and the centre of one artery, excluding the contribution of their respective radii to be $(X_0,0)$ (see Fig.~\ref{fig:smalld_coor}a). Here we assume the artery is centred on the $x$-axis. Then the coordinates local to this separation between the vessels can be written as 
\begin{equation}\label{eq:local_coord}
    X =X_0+\hat{x}d, \quad  Y = \hat{y}d^{1/2}.
\end{equation}
We approximate the circular cross-sections of the vessels by their leading-order parabolic form (see Fig.~\ref{fig:smalld_coor}b). For the vein  
\begin{equation}
    X = X_0 - \frac{d}{2}-\frac{Y^2}{2r_v}+... 
\end{equation}
while the cross-sections of the arteries are described by
\begin{equation}
    X = X_0 + \frac{d}{2}+\frac{Y^2}{2r_a}+...
\end{equation}

Then we rewrite \eqref{eq:2Dc} in local coordinates
\begin{equation}\label{eq:diff_smalld}
    \frac{1}{d^2}C_{\hat{x}\hat{x}} +\frac{1}{d}C_{\hat{y}\hat{y}}+\Omega^2
\left[ \frac{\hat{y}}{d}^2C_{\hat{x}\hat{x}}+\frac{X^2_0}{d}C_{\hat{y}\hat{y}}-\frac{X_0}{d}C_{\hat{x}}-\frac{2X_0\hat{y}}{d}C_{\hat{x}\hat{y}}+...\right] =0.
\end{equation}
Assuming that $\Omega$ is of order $\mathcal{O}(1)$, then solving \eqref{eq:diff_smalld} for $d \to 0$ which is $C_{\hat{x}\hat{x}}=0$ with boundary conditions 
\begin{subequations}
    \begin{align}
        C = 0 \quad\text{on}\quad \hat{x} &= \frac{1}{2}+\frac{\hat{y}^2}{2r_a}\\
        C = 1 \quad\text{on}\quad \hat{x} &= -\frac{1}{2}-\frac{\hat{y}^2}{2r_v}
    \end{align}
\end{subequations}
gives 
\begin{equation}
    C =\frac{1/2 +\hat{y}^2/(2r_a)-\hat{x}}{1+\hat{y}^2/(2r)}\,,\quad\text{where}\quad \frac{1}{r} = \frac{1}{r_a}+\frac{1}{r_v}\,.
\end{equation}

In order to calculate the flux between \chg{the vein and the nearby artery} we use \eqref{eq:fluxSt} with the local coordinates \eqref{eq:local_coord}:
\begin{equation}
    \int \left[\frac{C^2_{\hat{x}}}{d^2} + \frac{C^2_{\hat{y}}}{d} +  \frac{\Omega^2}{d}(\hat{y} C_{\hat{x}}-X_0C_{\hat{y}} +...)^2\;\right] d^{3/2} \d\hat{x} \d \hat{y}.
\end{equation}
For $d \to 0$, the leading order flux is
\begin{align}\label{eq:smalldN}
     \frac{1}{\sqrt{d}} \int C^2_{\hat{x}} \d\hat{x} \d\hat{y} &= \frac{1}{\sqrt{d}} \int_{-\infty}^{\infty} \d\hat{y} \int_{-\frac{1}{2}-\frac{\hat{y}^2}{2r_v}}^{\frac{1}{2}+\frac{\hat{y}^2}{2r_a}}  \frac{\d\hat{x}}{\left[1+\hat{y}^2/(2r)\right]^2} \nonumber\\
     &= \frac{1}{\sqrt{d}} \int_{-\infty}^{\infty} \frac{\d\hat{y}}{1+\hat{y}^2/(2r)}\\
    &= \pi\left( \frac{2r}{d} \right)^{1/2}. \nonumber     
\end{align} 
The leading order flux is of order $\mathcal{O}(d^{-1/2})$. The next corrections to the flux between the vessels are of order $\mathcal{O}(d^{1/2})$ and $\mathcal{O}(\Omega^2 d^{1/2})$ and the $\mathcal{O}(1)$ from the flux elsewhere. \chg{Thus,} when $d \to 0$, the flux between the vessels dominates over the flux elsewhere. \chg{Given the assumption that both arteries are the same distance $d$ from the vein,} the flux \chg{exchanged by} the vein need\chg{s} to be multipl\chg{ied} by \chg{a factor of two, as \eqref{eq:smalldN}} only takes into consideration the interaction between the vein and one of the arteries.

\section{Oxygen uptake process}\label{app:uptake}
\begin{figure}
    \centering
    \includegraphics[width=0.3\linewidth]{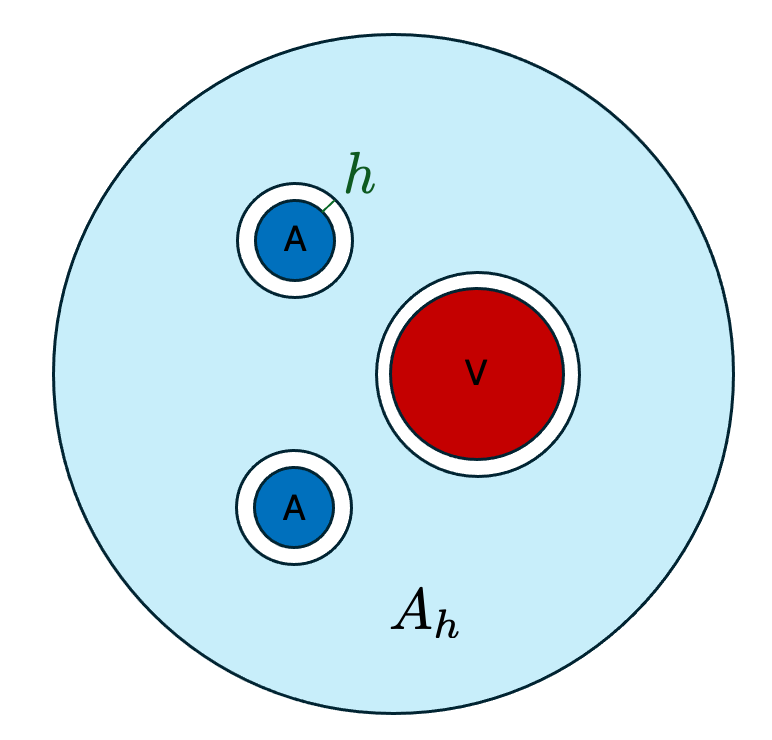}
    \caption{Diagram showing $A_h$ from \eqref{eq:N_u} the surface of the cord's tissue excluding a strip of thickness $h$ around each vessel.}
    \label{fig:Ah}
\end{figure}
\emph{Vasa vasorum} are smaller vessels that supply the walls of larger vessels with oxygen. In the human umbilical cord, there are no such vessels around the umbilical vessels \citep{Benirschke2012}. We hypothesize that the exchange between the vessels might help supply oxygen to the walls of the umbilical arteries and hence function as a `virtual' \emph{vasa vasorum}. In order to quantify the additional supply of oxygen to the umbilical arteries from the umbilical vein we calculate the solute flux taken up by the cord tissue excluding a thin strip around each vessel
\begin{equation}\label{eq:N_u}
    N_u = \alpha \int_{A_h} \left(c+ \frac{1}{\beta}\right)   \d A\,.
\end{equation}
Here $A_h$ defines the surface of the cord's tissue excluding a strip of thickness $h$ around each vessel (see Fig.~\ref{fig:Ah}).
By excluding this thin strip, we avoid unphysiologically large values due to an infinite supply of solute. The thickness of the strip $h = 0.05$ is comparable to the wall thickness of the vessels in \cite{Gayatri2017}.

Additionally, we define a baseline problem where only the arteries act as sources and exclude the contribution of the vein by imposing no flux at the vein--tissue interface. We scale the concentration with the fixed arterial concentration $c^*=c^*_a c$. Then 
\begin{subequations}\label{eq:up_can}
    \begin{align}
    \nabla^2 c &= \alpha\,c \;\text{ on }\; \chg{V}_t\, \text{,\; where }\; \alpha = \frac{q^*_0 (R^*_c)^2}{D_t^*}\,,\\
    \hat{\mathbf{n}} \cdot \nabla c &= 0 \;\;\;\,\text{ on }\; \chg{\partial V}_{\!0}\,,\\
     c &= 1 \;\;\;\,\text{ on }\; \chg{\partial V}_{\!i} \;\text{ for $i = 1,2$}\; \text{ and }\;  \hat{\mathbf{n}} \cdot \nabla c = 0 \;\text{ on }\; \chg{\partial V}_{\!3}\,.
\end{align} 
\end{subequations}
Similar to \eqref{eq:N_u}, we define the extravascular baseline uptake flux 
$N_b=\alpha \int_{A_h} c \d A$, which is calculated by excluding a thin strip of thickness $h$ around the arteries. 
To quantify the additional oxygen taken up by the tissue due to diffusive coupling of the UV and UAs, we evaluate
\begin{equation}\label{eq:Nup}
    N_s = \frac{N^*_u-N^*_b}{N^*_b} = \beta\,\frac{N_u}{N_b} - 1\,.
\end{equation}

\section{Validating the 2D approach with a 3D approach}
The 3D problem defined in \eqref{eq4:3Dc}, was solved using the software COMSOL Multiphysics 6.0. For computational purposes, we considered a section of the cord containing one coil, and then applied periodic boundary conditions on the top and bottom cross-sections to ensure that the two surfaces had equal concentration and flux. To calculate the flux leaving each vessel we used the \emph{Transport of Diluted Species} COMSOL module. The 2D problem \eqref{eq:2Dc} was solved using the \emph{General PDE} COMSOL module. We verified that the results for 2D and 3D approaches are consistent. To demonstrate this, firstly we compare flux predictions for different $\Omega$ in Table \ref{tab:valid}. The results agree to three decimal places, apart from $\Omega = 3.77$\chg{, which correspond to the relative difference of less than 0.5\%}.
\begin{table}
\centering
\begin{tabular}{lll}
\toprule
\textbf{Helicity} $\Omega$\hspace{2em} & \textbf{3D} $N$\hspace{2em} & \textbf{2D} $N$\\ \midrule
3.77  & 20.022  & 19.934  \\
1.885 & 13.276  & 13.276  \\
1.257 & 11.753  & 11.752  \\
0.942 & 11.152  & 11.152  \\
0.754 & 10.854  & 10.854  \\
0.628 & 10.686  & 10.686  \\ 
\bottomrule
\end{tabular}
\caption{\chg{The exchange} flux per unit length, $N$, for \chg{the} 2D problem \eqref{eq:2dflux} and 3D problem (\eqref{eq4:Ni} divided by pitch) for different \chg{values of} $\Omega$.}
\label{tab:valid}
\end{table}

\begin{figure}
    \centering
    \includegraphics[width=0.35\textwidth]{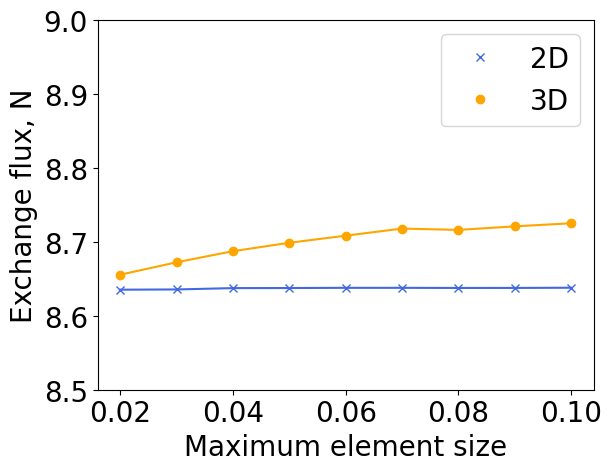}
    \caption{\chg{A comparison of the} 2D and 3D results for $\Omega = 3.77$ by varying the maximum element size.}
    \label{fig:mesh}
\end{figure}
A mesh convergence analysis was carried out for $\Omega = 3.77$ to explore how different meshing affects the flux calculation. The mesh was refined by decreasing the maximum element size from $0.1$ to $0.02$ while keeping the minimum size as $1\times10^{-4}$. Fig.~\ref{fig:mesh} shows how the flux converges by using increasingly finer mesh. The flux from the 2D calculation converges faster than the 3D results and the 3D calculation uses 300 times more elements than the 2D result. Results from the 3D and 2D computations agree within $1\%$ for a mesh with the maximum finite-element size of 0.02. Given its accuracy and computational efficiency, we adopt the 2D approach for all simulations.

\section{Image analysis of histology and ultrasound data}\label{app:xsec_extract}
\begin{figure}
    \centering
    \includegraphics[width=0.4\linewidth]{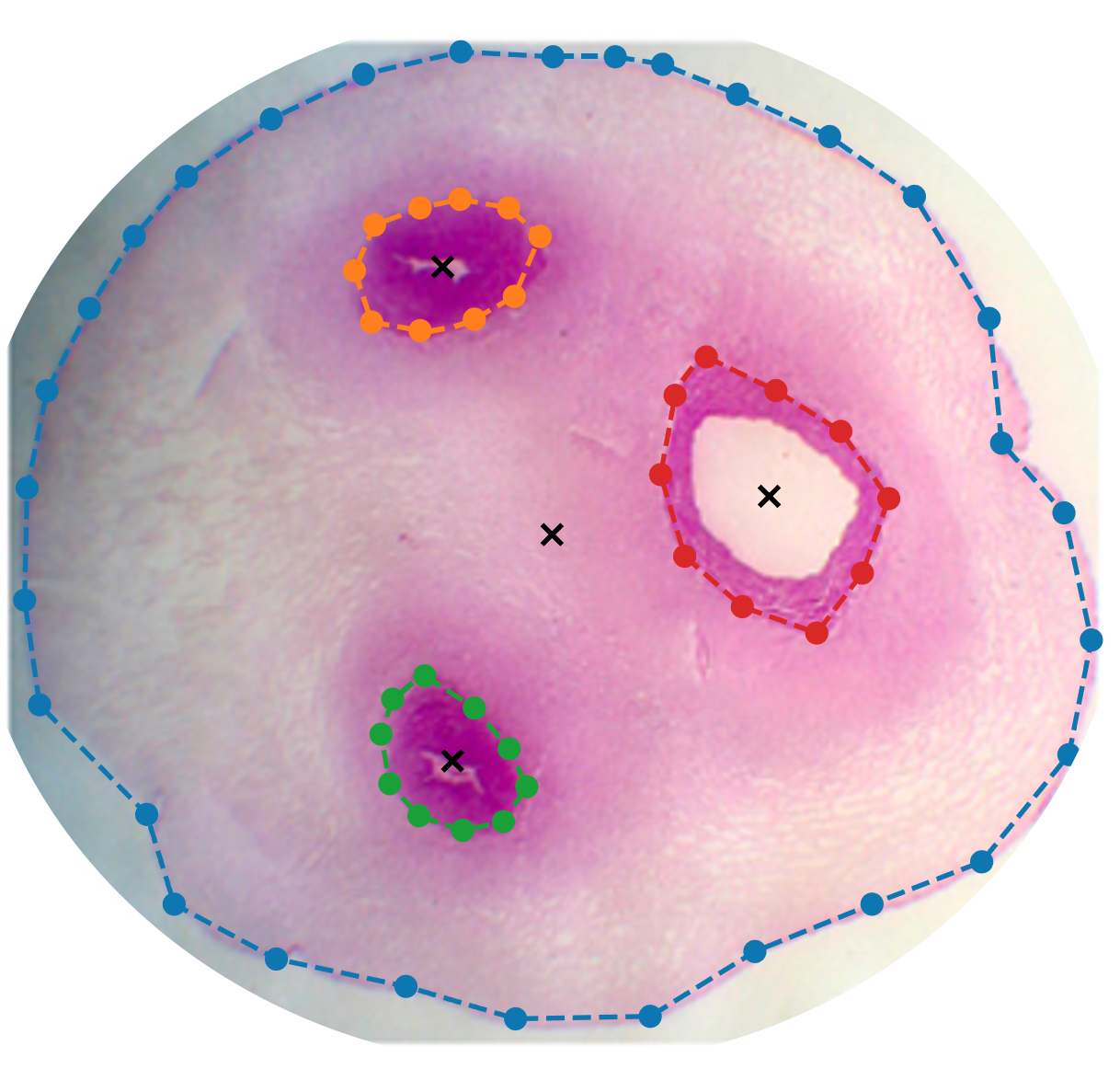}
    \caption{\chg{Example of the extracted outline of a cord and its vessels by interpolating the points, which are manually selected by the user. The black crosses show the centroids of the cord and the vessels calculated from the interpolated outlines. The image is adapted from \cite{Blanco2011} under CC BY-NC 4.0.}}
    \label{fig:interpo}
\end{figure}
The semi-automated Python tool \chgrev{(available at \url{https://doi.org/10.6084/m9.figshare.28382012})} allows us to extract the outline of the cord and each vessel \chg{by manually selecting points} on \chg{the loaded} image. The extracted outline is then interpolated using B-splines from which the centroids of the cord and the vessels are calculated \chg{(see Fig.~\ref{fig:interpo})}. The angles are measured as shown in Fig.~\ref{fig:geom_cord}(c). The angle $\theta_1$ is measured relative to the $x$-axis (the line passing through the centre of the cord and centre of the vein) assuming the vein is offset to the positive side of the $x$-axis. $\Delta\theta$ is measured as the angle between the lines connecting the centre of the vein and the centres of the arteries (see Fig.~\ref{fig:geom_cord}). \chg{To estimate the distance $d$ between the vein and the arteries we first find the line connecting the centroids of the vein and the artery. Then we find the points on the vessels' outlines that intersect this line. Finally, the distance $d$ between these two points is scaled by the radius of the cord.}

\begin{table}
\begin{tabular}{lllllll}
\toprule
$\theta_1$\hspace{2em} & $\Delta\theta$\hspace{1em} & \chg{$d$ \textbf{(Artery 1)}} & \chg{ $d$ \textbf{(Artery 2)}} &  \textbf{Gestational age}   & \textbf{Modality}\hspace{2em} & \textbf{Source}             \\ \midrule
135        & 75  & \chg{0.41} & \chg{0.45}    & 39 weeks          & Histology        & \cite{Blanco2011}       \\
154        & 91  & \chg{0.60} &  \chg{0.12}           & $37-40$ weeks       & Histology        & \cite{Thomas2020}     \\
159        & 47  &  \chg{0.70} &  \chg{0.30}            & $37 - 42$ weeks     & Histology        & \cite{Herzog2017}      \\
 164        & 51    &  \chg{0.41} &  \chg{0.41}          & $37 - 42$ weeks     & Histology        & \cite{BlancoElices2022} \\
165        & 42    &  \chg{0.40} &  \chg{0.49}          & $37 - 42$ weeks     & Histology        & \cite{BlancoElices2022} \\
167        & 44     &  \chg{0.43} &  \chg{0.65}        & $37 - 42$ weeks     & Histology        & \cite{BlancoElices2022} \\
165        & 41    &  \chg{0.48} &  \chg{0.66}         & $37 - 42$ weeks     & Histology        & \cite{BlancoElices2022}\\
165        & 53    &  \chg{0.38} &  \chg{0.40}         & $37 - 42$ weeks     & Histology        & \cite{BlancoElices2022} \\
162        & 52    &  \chg{0.54} &  \chg{0.54}         & $37 - 42$ weeks     & Histology        & \cite{BlancoElices2022} \\
164        & 40    &  \chg{0.29} &  \chg{0.29}         & $37 - 42$ weeks     & Histology        & \cite{Benirschke2012}   \\
151        & 45     &  \chg{0.19} &  \chg{0.62}        & $37 - 42$ weeks     & Histology        & \cite{Benirschke2012}  \\
145        & 50      &  \chg{0.29} &  \chg{0.30}       & 38 weeks          & Histology        & \cite{Kurakazu2019}    \\
146        & 41   &  \chg{0.59} &  \chg{0.38}          & 2nd$-$3rd trimester\hspace{1em} & Ultrasound       & \cite{Kurita2009}      \\
163        & 48     &  \chg{0.25} &  \chg{0.22}        & $37 - 42$ weeks     & Ultrasound       & \cite{HenanDh2016}     \\
\chg{160}   & \chg{66}     &  \chg{0.18} &  \chg{0.04}    & \chg{$12 - 14$ weeks}     & \chg{Ultrasound}  & \cite{Barbieri2008} \\
\bottomrule
\end{tabular}
\caption{Details on histology and ultrasound data.}
\label{tab:histdata}
\end{table}

\section{Solute exchange in an asymmetric configuration and effects of chirality}
\begin{figure}
    \centering
    \includegraphics[width=0.9\textwidth]{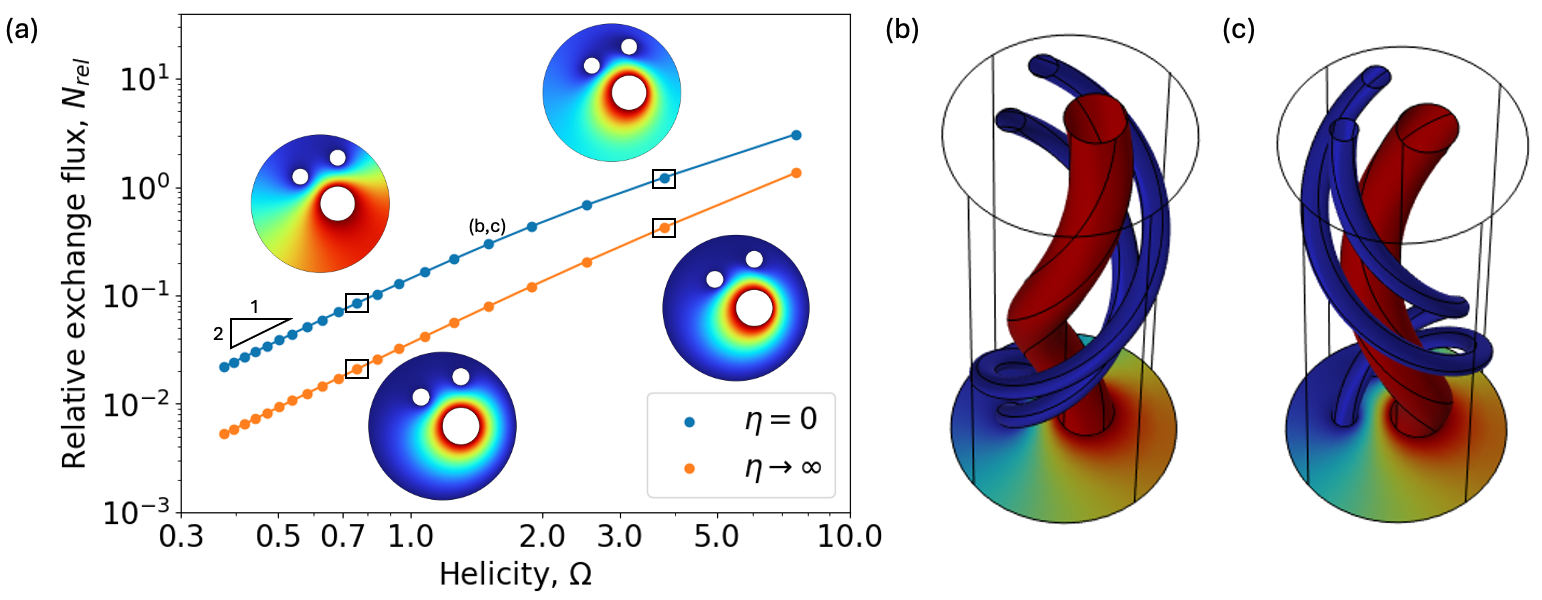}
    \caption{(a) Exchange flux for an asymmetric configuration with $\theta_1=\pi/2$ and $\theta_2=0.8\pi$. Concentration fields for $\Omega=0.75$ and $\Omega=3.77$. 3D configurations for $\Omega = 1.5$ with (b) a left-handed helix and (c) a right-handed helix. 
    Other geometric parameters were fixed: $R_v = 0.25$, $r_v = 0.25$, $r_a = 0.12$, $d = 0.3$ and $\eta = 0$. \chg{The triangle in (a) shows the relative fold-change in the variables.}}
    \label{fig:asym}
\end{figure}

Fig.~\ref{fig:asym}(a) displays how $N_{rel}$ changes with helicity for a configuration with asymmetric arteries. The same trend in relative flux as in the symmetrical case (Fig.~2, main text) can be seen but breaking the symmetry introduces distortions in the concentration fields.  Additionally, when the cord boundary acts as a sink ($\eta\to\infty$), the effects of helicity are weaker, as shown by the concentration fields. When $\eta=0$, as helicity increases the concentration field becomes more uniform but when $\eta\to\infty$ there is less relative change in the concentration field as helicity increases. Fig.~\ref{fig:asym}(b) confirms that configurations with opposite handedness have identical concentration fields showing that chirality has no effect on solute transfer. Table \ref{tab:histdata} shows the values of $\theta_1$ and $\Delta\theta$ and the corresponding source.

\section{Estimation of the Damk\"{o}hler number}
\begin{table}
\centering
\begin{tabular}{lll}
\toprule
\textbf{Parameter}     & \textbf{Value}                     & \textbf{Source} \\ \midrule
Oxygen diffusivity, $D^*_\text{ox}$   & $2 \times 10^{-5}$ \si{cm^2/s}   &    \cite{Jordan1956}      \\
\chg{Thermal} diffusivity, $D^*_\text{th}$ & $1.4 \times 10^{-3}$ \si{cm^2/s} &   \cite{Ponder1962}\\
\chg{UV} flow rate, $Q^*_v$        & $3$ \si{cm^3/s}           &    \cite{Lees1999}    \\ 
Oxygen advective factor, $B$        & $140$          & \cite{Kaesinger1981}   \\
Heat advective factor, $B$        & $1$     &    \\
Oxygen concentration in \chg{UA}, $c^*_a$        & $16-28$ \si{mmHg}     &  \cite{Nye2018}  \\ 
Oxygen concentration in \chg{UV}, $c^*_v$        & $19-43$ \si{mmHg}     &  \cite{Nye2018}   \\
\chg{Oxygen} uptake rate, $q^*_0\,\chg{=q^*_\text{max}/c^*_{50}}$        & $10^{\chg{-2}}-1$ \si{s^{-1}}    & \chg{\cite{Erlich19-metabolism}}\\ 
\chg{Oxygen uptake parameter, $\alpha=q^*_0\,(R^*_c)^2/D^*_\text{ox}$}   & \chg{$\sim 10^{3}-10^{5}$}  & \chg{Estimated}\\
\chg{Relative oxygen concentration, $\beta=(c^*_v-c^*_a)/c^*_a$}         & \chg{$\approx 0.2 - 1.5$}  & \chg{Estimated}\\ 
\bottomrule
\end{tabular}
\caption{Key physico-chemical parameters of exchange in the umbilical cord.}
\label{tab4:D}
\end{table}

After obtaining the flux per unit length, $N$, from the two-dimensional model we can characterise solute exchange and its clinical significance by estimating the ratio of diffusion rate and flow rate for the helical umbilical vein. In this section, we focus on oxygen and heat exchange.

We define the flux ratio as the Damk\"{o}hler number
\begin{equation}
    \Da = \frac{\text{Tissue diffusion rate}}{\text{Umbilical flow rate}} = \frac{N D^*_{\chg{t}} L^*}{BQ^*_v}.
\end{equation}
The results of Section 3(a) in the main text provide the two-dimensional flux per unit length $N\approx \mathcal{O}(10)$, defined by \eqref{eq:2dflux}. We take $L^* = 50$\,cm as the length of the umbilical cord \citep{Balkawade2012} and $Q^*_v = 3$\,\si{cm^3/s} as the umbilical vein flow rate \citep{Lees1999}. The parameter $B \approx 140$ models linearised oxygen--haemoglobin binding kinetics in the fetal blood that boosts its advective capacity \citep{Kaesinger1981} and $D^*_{\chg{t}}$ is the diffusivity \chg{of solute or heat in tissue}  (see Table \ref{tab4:D}). The resulting Damk\"{o}hler number estimate for oxygen is of the order of $10^{-5}$. This suggests that oxygen transport is diffusion-limited, where the flow rate dominates over the diffusion rate. Thus there is virtually no exchange in the umbilical vessels.

On the other hand, heat exchange is characterised by the lack of advective facilitation ($B=1$) and a more rapid thermal diffusivity $D^{\chgrev{*}}_\text{th} \sim 10^{3}$\,cm$^2$/s (\emph{vs.} $D^{\chgrev{*}}_\text{ox} \sim 10^{5}$\,{cm$^2$/s} for oxygen; see Table~\ref{tab4:D}). The corresponding value of the Damk\"{o}hler number is thus of the order of $10^{-1}$, indicating the possibility of heat exchange between the umbilical vessels.

\newpage
\bibliography{references.bib}
\bibliographystyle{RS} %